\documentclass[twocolumn,showpacs,aps,pra,superscriptaddress]{revtex4-2}
\include{def}
\usepackage{amsmath}
\usepackage{lineno}
\usepackage{xcolor}
\usepackage{xcolor}
\colorlet{RED}{red}
\usepackage{longtable}
\usepackage{graphicx}
\usepackage{dcolumn}
\usepackage{bm}

\begin{document}

\setlength\LTcapwidth{\linewidth}
\title{Features of hadronic and deconfined matter from AGS to LHC energies}
\author{M. Petrovici}
\affiliation{National Institute for Physics and Nuclear Engineering - IFIN-HH, P.O. Box MG-6,\\
Hadron Physics Department\\
Bucharest-Magurele, Romania}
\author{A. Pop} 
\affiliation{National Institute for Physics and Nuclear Engineering - IFIN-HH, P.O. Box MG-6,\\
Hadron Physics Department\\
Bucharest-Magurele, Romania}
\date{\today}
\begin{abstract}
Previous extensive studies on the dependence
of the average transverse momentum, its slope as a function of
the hadron mass and the average transverse expansion on
the particle multiplicity per unit rapidity and  unit transverse 
overlap area of the colliding partners
are extended to the ratio of the energy density
to the entropy density.

The 
behaviour of the ratio between the average transverse momentum and the square root of the particle multiplicity per unit rapidity and unit transverse overlap area $\langle p_T \rangle/\sqrt{\langle dN/dy\rangle/S_{\perp}}$
as a function of collision energy for a given centrality or
as a function of centrality for a given collision energy supports the predictions
of CGC and percolation based approaches. 
The dependence of
the ratio of the energy density $\langle dE_{T}/dy\rangle/S_{\perp}$ to the 
entropy density $\langle dN/dy\rangle/S_{\perp}$ at different collision centralities
for A-A collisions from AGS, SPS, RHIC and  LHC
energies is presented. 
The trend of this ratio towards a plateau at the highest RHIC energies followed by a steep rise at LHC energies 
is in agreement with theoretical predictions made 40 years ago that 
indicate this behaviour as a signature of a phase transition.
This pattern strongly depends on the collision geometry, converging towards the dependence 
that characterizes the pp minimum bias (MB) collisions for the most peripheral A-A collisions.
Expected similarities between pp and 
Pb-Pb collisions at LHC energies are confirmed.  
\end{abstract}
\maketitle

\section{Introduction}
%\linenumbers

The unprecedented amount of experimental information obtained at
AGS, SPS, RHIC up to the highest 
energies from LHC supports the theoretical predictions made
about 50 years ago on the possibility to produce very hot
and dense matter in heavy ion collisions \cite{Cha}. 
Using QCD asymptotic 
freedom properties, a transition from the hadronic phase to a high
density "quark soup" \cite{Coll, Cab} or "quark-gluon plasma"
\cite{Shu} following the "quark liberation" idea \cite{Cab}
is expected. The first estimates of the transition from a gas of free 
nucleons to hadronic matter and subsequently to deconfined matter as
a function of density were done within the percolation approach 
\cite{Bay,Cel}. 
Studying the production mechanism and properties of tiny pieces of deconfined matter 
as a function of temperature and density is a difficult task to the extent that the highly 
inhomogeneous initial state, finite size effects, and violent dynamical evolution 
must be taken into account to obtain a  unique explanation of the 
experimental observations.
That is why a clean and unambiguous experimental signature of deconfinement using collisions 
with heavy ions is rather difficult to be established.
Theoretical approaches, which combine sound hypotheses for different snapshots 
of the formation and evolution of the system produced in heavy ion collisions 
starting with the initial phase described in the Color Glass Condensate (CGC) approach \cite{McL1, McL2, McL3, Ian},
 followed by the process of equilibration described on the basis of the QCD kinetic theory method 
 and the hydrodynamic expansion were recently developed \cite{Maz}.
 The correlation between the multiplicity of particles and the average transverse momentum ($\langle p_T \rangle$)
 has been proposed as a signature of the transition from hadronic to deconfined matter \cite{Van}.
 Such phenomenological models predict an increase followed by a plateau 
 of  $\langle p_T \rangle$ as a function of entropy density, caused by the 
 mixed phase corresponding to the transition from pure hadronic matter to deconfined matter.
 At even larger entropy density
the $\langle p_T \rangle$ starts to increase again. 
Following this idea and existing information on the average transverse mass ($\langle m_T \rangle$) 
of charged hadrons identified in A-A collisions at AGS, SPS, and RHIC, 
a saturation of $\langle m_T \rangle$ as a function of the charged particle multiplicity
per unit rapidity for the most central collisions  was evidenced \cite{Moh}.
In this study the $\langle m_T \rangle$ was estimated 
using an exponential parameterization of the experimental 
$p_T$ spectra. 
Recently, including more information obtained at RHIC in the Beam Energy Scan (BES) program for Au-Au collisions 
and LHC data for the Pb-Pb collision at $\sqrt{s_{NN}}$ = 2.76 TeV \cite{Adam} the correlation ($\langle m_T \rangle-m_{0})$-ln($\sqrt{s_{NN}}$) 
for the most central collisions was studied for identified hadrons under the hypothesis 
that ($\langle m_T \rangle-m_{0})$ can be an approximation of the system temperature  while 
ln ($\sqrt{s_{NN}}$)$\sim \langle dN/dy\rangle$ of its entropy. $m_{0}$ is the rest mass of the particle.
 A dependence on mass was clearly observed.
The saturation range in 
ln($\sqrt{s_{NN}}$) decreases and ($\langle m_T \rangle-m_{0}$) 
increases again more violently going from pions to protons. 
It is worth mentioning that collective expansion and suppression phenomena 
can contribute in such a correlation to the observed trend.
A similar prediction was obtained by including the relativistic hydrodynamic evolution 
of the quark-gluon plasma \cite{Bla1, Bla2}. 
In these works it was suggested that the correlation between the ratio of the energy density to entropy density 
is also sensitive to the transition from hadronic to deconfined matter.
The energy density per unit transverse overlap area can be estimated using the Bjorken invariance, 
while the entropy density per unit transverse overlap area, assuming an isentropic expansion, 
is proportional to the particle multiplicity per unit rapidity and unit transverse overlap area.
A similar correlation, using the charged particle multiplicity  and the transverse energy per unit pseudorapidity 
($\langle dE_T/d\eta \rangle/\langle dN_{ch}/d\eta \rangle)$ 
for the most central collisions, with the assumption that ln($\sqrt{s_{NN}}$)$\sim dN_{ch}/d\eta$  and a rather 
small dependence of the transverse overlap area ($S_{\perp}$) on the collision energy, 
was plotted using experimental information from SIS18 up to the highest energy from RHIC \cite{Adl},
 the results from the LHC for the Pb-Pb collision at $\sqrt{s_{NN}}$ =  2.76 TeV being included
  once the data became available \cite{Adam1}.
As predicted by the above mentioned models a plateau at the highest energies at RHIC  
and a strong increase at the LHC energy were evidenced.
The $\langle dE_T/d\eta \rangle/\langle dN_{ch}/d\eta \rangle$-$\sqrt{s_{NN}}$
correlations for the most central collisions were shown for 
data measured by the PHENIX Collaboration at RHIC energies \cite{PHENIX1}.
The $\langle dE_T/dy \rangle/\langle dN_{ch}/dy \rangle$-$\sqrt{s_{NN}}$
correlations for different centralities were represented for 
data measured by the STAR Collaboration at BES energies \cite{Bis}.
Theoretical predictions show that the
energy over entropy versus entropy correlation 
is more sensitive to the equation of state than the 
average transverse mass versus multiplicity density one \cite{Monn}.

In the present paper, the results for the
$\langle dE_T/dy \rangle/\langle dN/dy \rangle$-$\langle dN/dy \rangle/S_{\perp}$ 
correlation as a function of centrality in A-A collisions at different energies and a
comparison with pp minimum bias and pp as a function of
$\langle dN/dy \rangle/S_{\perp}$ at LHC energies are presented. 
Results based on published data obtained in Au-Au collisions at AGS and 
RHIC and Pb-Pb collisions at SPS and LHC, in terms of 
$\langle dE_T/dy \rangle/\langle dN/dy \rangle$-$\langle dN/dy \rangle/S_{\perp}$ 
for the most central collisions are presented in Section II. 
Section III is dedicated to similar studies as a function of centrality for
RHIC and LHC energies. 
In Section IV the core contribution in such correlations is presented. The
dependence on $\langle dN/dy \rangle$/$S_{\perp}$ of the Bjorken energy density times the interaction time for 
central collisions and as a function of centrality is presented in 
Section V.
The comparison of A-A with MB pp collisions at the same energies and for different
charged particle multiplicities  at LHC energies is discussed in
Section VI.
 Conclusions are presented in Section VII.

\section{$\langle dE_T/dy \rangle/\langle dN/dy \rangle - \langle dN/dy 
\rangle/S_{\perp}$ correlation for central A-A collisions}

\begin{figure*}
\includegraphics[width=0.48\linewidth]{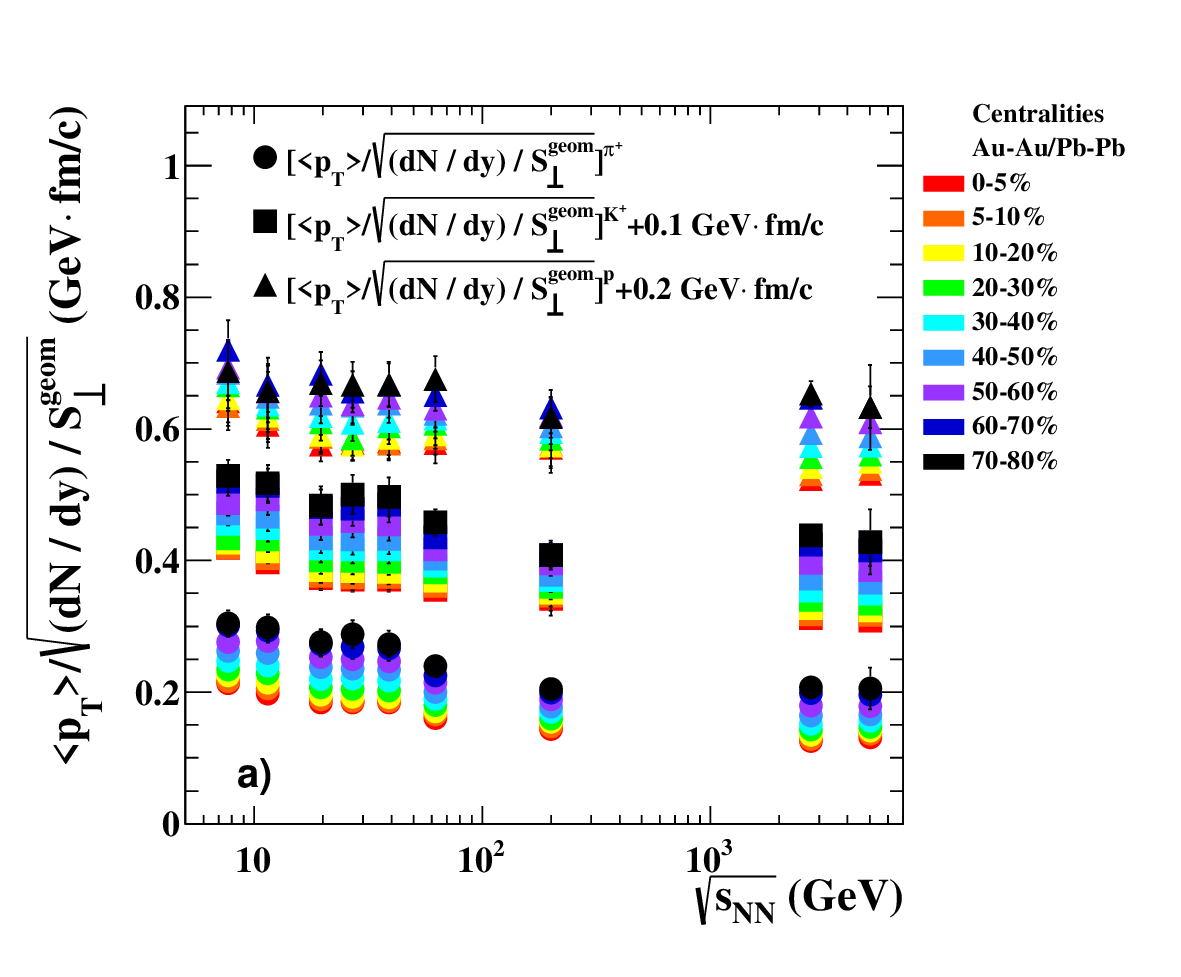}
\includegraphics[width=0.50\linewidth]{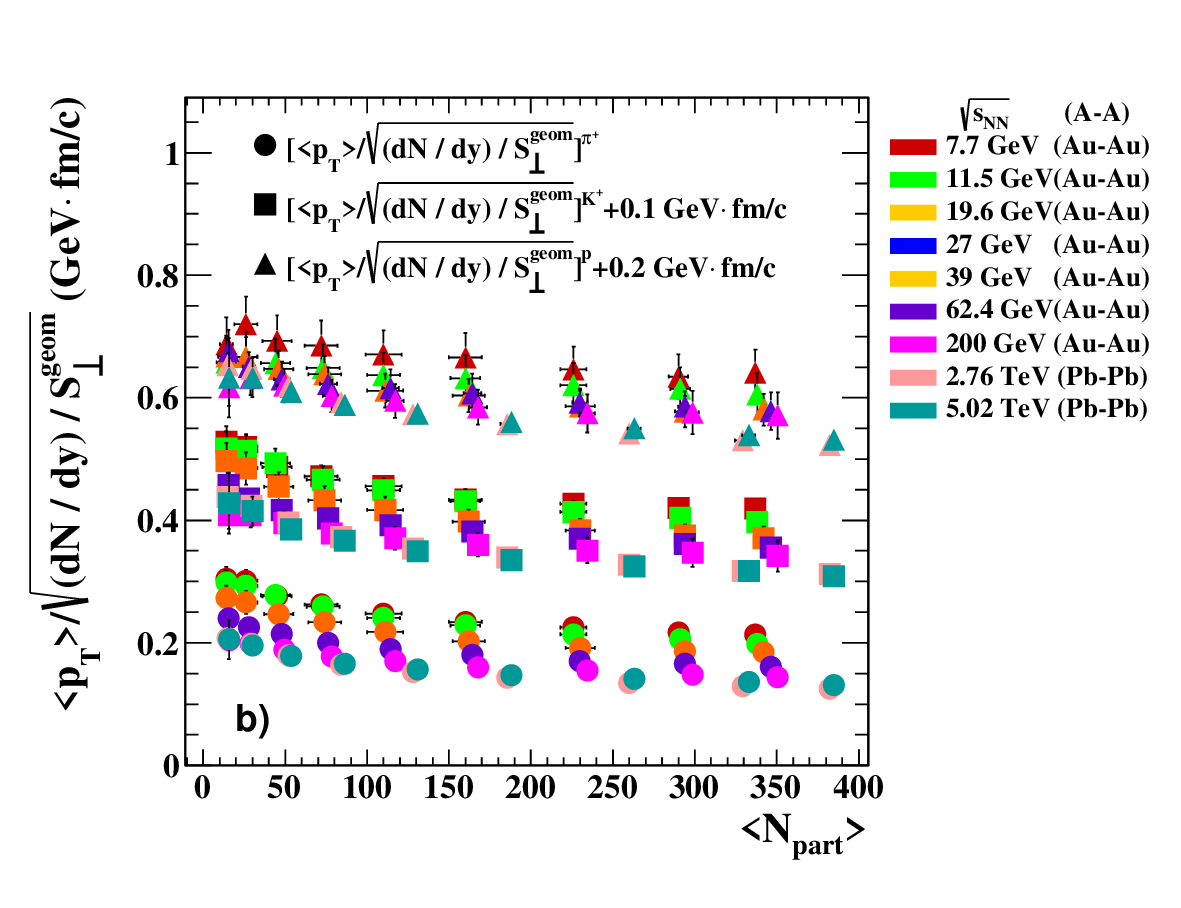}
\caption{$\langle p_T\rangle/\sqrt{(dN/dy)/S_{\perp}^{geom}}$ dependence on:
a) collision energy for different centralities and 
b) $\langle N_{part} \rangle$ for different collision energies.}
\label{fig-1}
\end{figure*}

In our previous
studies \cite{Pet1,Pet2,Pet3} it was shown that the particle
multiplicity per unit rapidity and unit transverse overlap area, a good estimate of
the entropy density, turns out to be a scaling observable
which governs the behaviour of the average transverse expansion 
for identified
charged hadrons and hyperons, independent
on the size of the colliding systems and even for pp collisions at 
LHC energies. For collision energies higher than $\sqrt{s_{NN}}$=39 GeV the
scaling 
was also evident for the slope of $\langle p_T \rangle$ as a function of 
mass.
While such scaling  also occurs below $\sqrt{s_{NN}}$=39 GeV up to the 
mid-central collisions, a slight
deviation of $\sim$10\% relative to the higher energies is observed 
towards central collisions.
As far as concerns $\langle p_T \rangle$, a very good scaling is observed for the 
identified 
charged hadrons at the RHIC energies. The linear dependence of 
$\langle p_T \rangle$ on $\sqrt{\langle dN/dy \rangle/S_{\perp}^{geom}}$ holds also for LHC 
energies, with a slightly different offset relative to the RHIC energies and a
 tendency towards saturation at
the most central collisions. The geometrical transverse overlap area ($S_{\perp}^{geom}$) of the two colliding nuclei for a given
incident energy and centrality was estimated on the basis of a Glauber Monte-Carlo (MC) approach \cite{Glauber55,Franco66,Miller07,Glis} 
as explained in \cite{Pet1}  where its values were compiled. They are used in the estimates made in the present paper.
Using also the experimental $\langle p_T\rangle$ values for positive pions, kaons and 
protons and $\langle dN/dy\rangle$  compiled in \cite{Pet1}, in Fig.~\ref{fig-1}a is represented the ratio 
$\langle p_T\rangle/\sqrt{\langle dN/dy\rangle/S_{\perp}^{geom}}$ as a function of 
collision 
energy for different centralities while in Fig.~\ref{fig-1}b as a function of 
centrality ($\langle N_{part} \rangle$)
for different collision energies.
As can be seen in Fig.~\ref{fig-1},
the ratio between the average transverse momentum  and the square root of 
the hadron 
multiplicity per unit rapidity and unit transverse overlap area of the colliding nuclei
decreases towards central collisions and higher energies, 
thus supporting the predictions based on colour fields description 
of the small x degrees of freedom \cite{McL1, Ian,Ianc1,*Ianc2,KLN1,*KLN2,*KLN3,*KLN4,*KLN5,*KLN6}, local parton-hadron 
duality picture (LPHD) \cite{LPHD} and 
dimensionality argument  \cite{Lev1, Lap1}.

A similar behaviour is predicted by the
string percolation scenario, inspired by CGC
\cite{Bir, Dias}. 
 \noindent
In a string percolation approach \cite{Dias}:
 \begin{equation}\label{eq1}
\frac{dN}{dy}=F(\eta)\bar{N}^s\mu_{0}
\end{equation}
\begin{equation}\label{eq2}
\langle p^2_T\rangle = \langle p^2_T\rangle_1/F(\eta)
\end{equation}
The colour reduction factor:
\begin{equation}\label{eq3}
F(\eta)\equiv\sqrt{\frac{1-e^{-\eta}}{\eta}}
\end{equation}
\noindent
where $\eta$, the 2-dimensional transverse density of strings, is given by $\eta\equiv(r_0/R)^2\bar{N}^s$, $r_0$ being 
the string radius and R the radius of the transverse overlap region. 
 $\bar{N}^s$ is the average 
number of strings, $\mu_{0}$  and $\langle p^2_T\rangle_1$ are the average particle density and respectively the average transverse 
momentum squared of a single string. 
Therefore, $\sqrt{\langle p^2_T\rangle}$/$\sqrt{\langle dN/dy\rangle/S_{\perp}}$$\sim 1/\sqrt{(1-e^{-\eta})}$ is decreasing with centrality for a given collision energy and
with collision energy for a given centrality.
The effects of the strong longitudinal colour fields on the identified charged 
hadrons and hyperons $p_T$ distributions in pp collisions at 
$\sqrt{s}$=7 TeV have also been investigated within the framework of the 
HIJING/B$\bar{B}$ v2.0 model \cite{Pet5}. The experimental ratios 
of the $p_T$ distributions at different charged particle multiplicities
to the one corresponding to the MB pp collision, each of them normalized to 
the corresponding charged particle density, were well reproduced using
an increased strength of the colour field, characterized by the effective 
values of the string tension, from low to high charged 
particle multiplicity. 
These results pointed out the necessity of 
introducing 
a multiplicity (or energy density) dependence for the effective value of 
the string tension. They also show that at the LHC energies the 
global features of the interactions are mainly determined by the properties
of the initial chromoelectric flux tubes, the system size playing a
minor role.
A rough estimate of the 
gluon number density and occupation number in the early stage of 
relativistic heavy ion collisions, following \cite{Muel}, gives values
by 2.3 (2.9) higher at $\sqrt{s_{NN}}$=2.76 TeV (5.02 TeV) Pb-Pb central collisions
relative  
to Au-Au central collisions at $\sqrt{s_{NN}}$=200 GeV.
For the largest charged particle multiplicity in pp collisions at 7 TeV
the values are a bit larger than those corresponding to Pb-Pb 
central collisions at
$\sqrt{s_{NN}}$=5.02 TeV.
The parton density evolution as a function of x and
 $Q^2$, addressed more than 35 years
ago \cite{GRIB1} was experimentally confirmed at HERA 
\cite{HERA1}. The rise of the structure function at 
low x is still visible at small values of $Q^2$ \cite{HERA2, HERA3} where the
perturbative QCD does not work anymore. Therefore, rich and short-lived partonic cascades
become visible as the collision energy increases, the string density and the percolation probability increasing too.
The approaches mentioned
above, predict an increase of the average transverse momentum at the 
same particle multiplicity per unit rapidity and unit transverse overlap area
with collision energy and system size independence of the global features evidenced at
LHC energies, the properties of the initial color electromagnetic 
flux tubes playing the main role. 

\begin{figure} []
\includegraphics[width=0.85\linewidth]{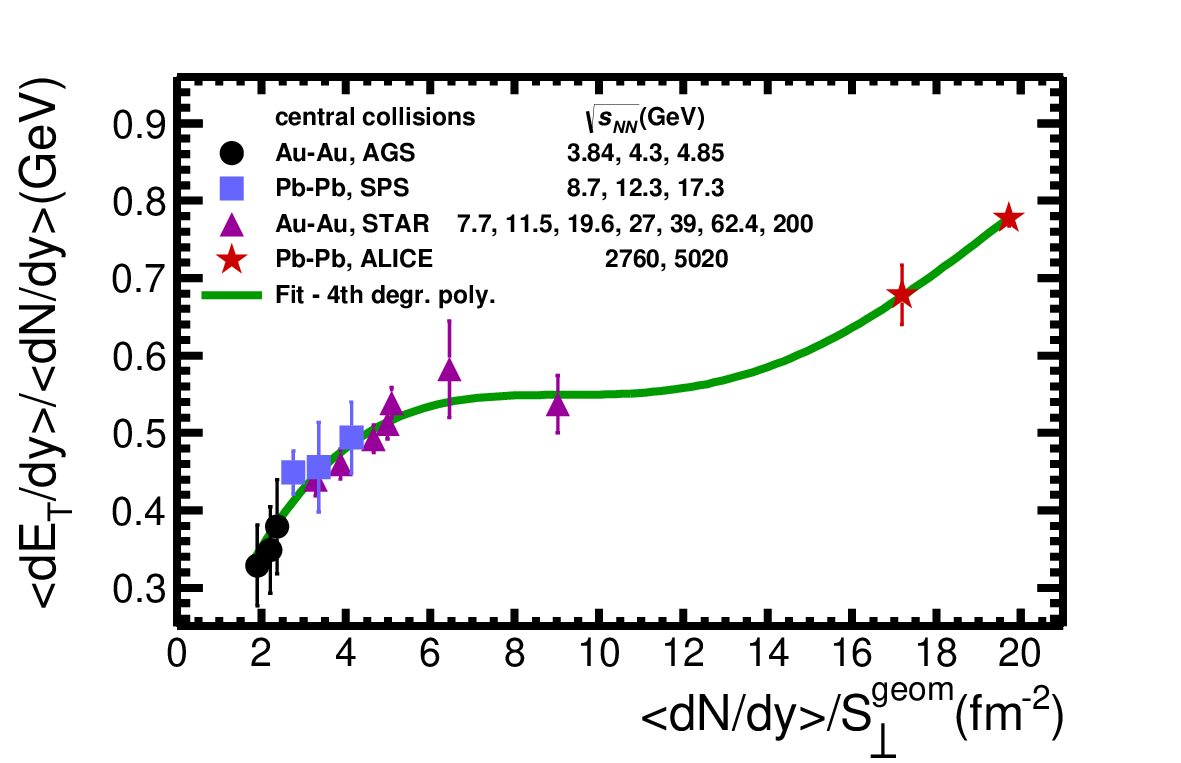}
\vspace{-0.5cm}
\includegraphics[width=0.85\linewidth]{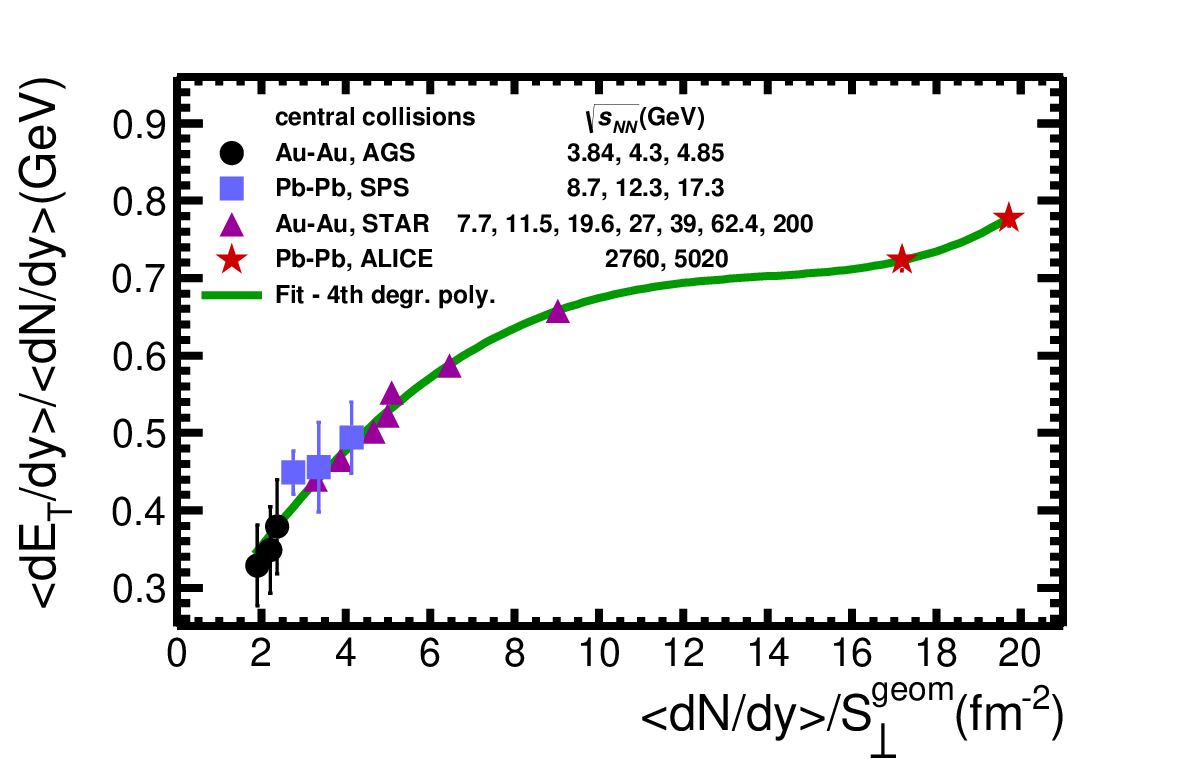}
%\vspace{-1.5cm}
\caption{The $\langle dE_T/dy \rangle/\langle dN/dy \rangle)$-$\langle dN/dy \rangle/
S_{\perp}$ correlation for the most central A-A collisions. Upper plot: case I, bottom plot: case II.
Figs. 2a and 2b are quite different at the highest energies from RHIC. 
This is mainly due to the 
difference at $\sqrt{s_{NN}}$=200 GeV between the transverse energy measured in \cite{STAR2} (case I) and 
the one estimated in this paper from the existing particle spectra (case II). The difference is about 20\%,
 the largest compared to the other energies, while the $\langle dN/dy \rangle$ value is the same.
For the AGS and SPS energies, since only one set of data is available, as 
cited in the text \cite{Adl}, the two cases coincide.
The continuous
 lines, used to guide the eye, represent the results of a
 $4^{th}$ degree polynomial fits of the experimental points.}
\label{fig-2}
\end{figure}

The $\langle dE_T/dy \rangle/\langle dN/dy \rangle$-$\langle dN/dy \rangle/S_{\perp}$ correlation for the most central A-A collisions using 
the experimental information collected over the years starting from AGS up to the LHC
energies,  is presented in Fig.~\ref{fig-2}. In the case of RHIC and LHC, experimental data from the STAR and ALICE 
collaborations were used mainly due to the availability in terms of measured spectra  for light-flavor hadrons. 
The transverse energy per unit rapidity was considered in two ways. In case I (upper plot of Fig.~\ref{fig-2}) published measured values 
were used 
when they were available, while in case II (bottom plot of Fig.~\ref{fig-2}) an approximation similar to that from \cite{STAR1} was applied for the  $\langle dE_T/dy \rangle$ calculation.
For the AGS and SPS energies the  $\langle dE_T/d\eta \rangle$ values and the scaling factor 
for $\eta$ to y transition at midrapidity were taken from \cite{Adl} for all cases.  
For the BES energies the  STAR data from \cite{Bis} were used in case I. It should be noted that these were obtained  from the transverse momentum spectra measured up to $\Lambda$ particles.
For  higher energies, STAR  data  regarding $\langle dE_T/dy \rangle$ reported for  
Au-Au at $\sqrt{s_{NN}}$ = 62.4 GeV in \cite{Sahoo1}  and \cite{Sahoo2} 
  and respectively at
  $\sqrt{s_{NN}}$ = 200 GeV in
  \cite{STAR2} as well as the results of the transverse energy measurement for Pb-Pb at $\sqrt{s_{NN}}$ = 2.76 
  TeV from \cite{Adam1} were considered.

In case II, the full formula used to approximate  $dE_{T}/dy$ was:
\begin{widetext}
%\begin{center}
\begin{equation}\label{eq4}
\frac{dE_T}{dy}\simeq \frac{3}{2}(\langle m_T\rangle \frac{d N}{dy})^{(\pi^+ + 
\pi^-)}+ 
2(\langle m_T\rangle\frac{dN}{dy})^{(p+\bar{p}, \Xi^- +\bar{\Xi}^+)}  +
(\langle m_T\rangle\frac{dN}{dy})^{(K^++K^-, \Lambda + 
\bar{\Lambda}, \Omega^- + \bar{\Omega}^+)}+
2(\langle m_T\rangle\frac{dN}{dy})^{K^0_{S}}+
2(\langle m_T\rangle\frac{dN}{dy})^{(\Sigma^{+}+\Sigma^{-})}
\end{equation} 
%\end{center}
\end{widetext}
When the particle spectra were available, the average transverse mass was calculated. For Pb-Pb at $\sqrt{s_{NN}}$ = 5.02 TeV in the case of strange and multi-strange particles for which the transverse momentum distributions have not yet been published, the approximation 
$\langle m_T \rangle=\sqrt{\langle p_{T}^{2} \rangle+m_{0}^{2}}$ was used. In accordance with the convention made in calorimetric measurements, $\langle m_{T} \rangle$ was replaced by 
$\langle m_{T} \rangle-m_{0}$ for baryons and $\langle m_{T} \rangle+m_{0}$ for antibaryons.

In both cases the particle multiplicity per unit rapidity, dN/dy was estimated according to:
\begin{widetext}
\begin{equation}\label{eq5}
\frac{dN}{dy}\simeq \frac{3}{2}\frac{dN}{dy}^{(\pi^+ + \pi^-)}+ 
2\frac{dN}{dy}^{(p+\bar{p}, \Xi^- +\bar{\Xi}^+)}  +
\frac{dN}{dy}^{(K^++K^-, \Lambda + \bar{\Lambda}, \Omega^- + \bar{\Omega}^+)} + 2\frac{dN}{dy}^{K^{0}_{S}} +
2\frac{dN}{dy}^{(\Sigma^{+}+\Sigma^{-})}
\end{equation}
\end{widetext}
The experimental data in terms of transverse momentum spectra, yields and average transverse momenta were taken from: \cite{Chatt} (AGS and SPS), \cite {Adam,STAR4} (Au-Au, BES), \cite{STAR1,STAR5} (Au-Au at $\sqrt{s_{NN}}$ = 62.4 GeV), \cite{STAR1,STAR6,STAR7} (Au-Au at $\sqrt{s_{NN}}$ = 200 GeV), \cite{ALICE2,ALICE3,ALICE41,*ALICE42} (Pb-Pb at $\sqrt{s_{NN}}$ = 2.76 TeV) and \cite{ALICE5,ALICE6,ALICE7,ALICE8} (Pb-Pb at $\sqrt{s_{NN}}$ = 5.02 TeV).
Eqs. \ref{eq4} and \ref{eq5} were applied so that the same particles and 
anti-particles to be taken into consideration in both 
formulas depending, 
from system to system and case to case, on the most complete published 
experimental information. With the increase of the collision energy, the 
contribution of strange and multi-strange particles in the formulas becomes 
more and more important. Thus, for Au-Au at $\sqrt{s_{NN}}$ = 62.4 and $
\sqrt{s_{NN}}$ = 200 GeV and Pb-Pb at both LHC energies, the contribution 
of 
$\Sigma$ particles was also considered as explained in \cite{Pal,Adam1}.
In Fig.~\ref{fig-2}, $\langle dE_T/dy\rangle/\langle dN/dy\rangle$ as a 
function of  $\langle dN/dy \rangle/S_{\perp}$ for the most central 
collisions is shown.
\begin{figure}
\includegraphics[width=0.8\linewidth]{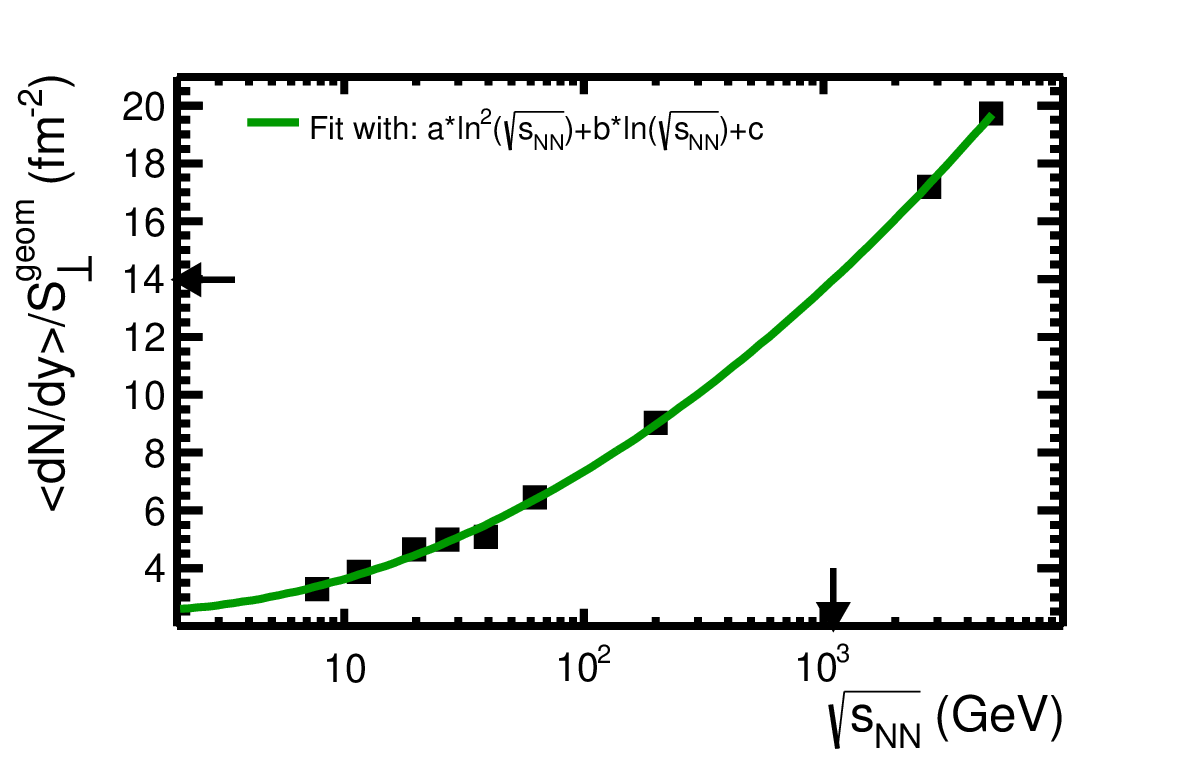}
\caption{$<dN/dy>/S_{\perp}^{geom}$ as a function of collision energy. 
The collision energy values correspond to the 
$\langle dN/dy \rangle/ S_{\perp}^{geom}$ values from Fig.2. The line is the result 
of the fit with a second order 
polynomial in $ln(\sqrt{s_{NN}})$. The two arrows that point to the x and y 
axes indicate the approximate value of the collision energy and $<dN/dy>/S^{geom}_{\perp}$, 
respectively, at the inflection point in case II.}
\label{fig-3} 
\end{figure}
The continuous lines in the figures, used to guide the eye, are the results 
of  the fits of experimental points using a $4^{th}$ degree polynomial.  
$\langle dE_T/dy \rangle/\langle dN/dy\rangle$ increases with 
$\langle dN/dy \rangle/S_{\perp}$ from AGS to SPS
and RHIC energies, up to $\sqrt{s_{NN}}$=39 GeV. 
Above $\sqrt{s_{NN}}$=39 GeV the slope changes and a tendency towards 
saturation is evident.
 Even if the gap in the collision energy between 
$\sqrt{s_{NN}}$=200 GeV, the highest one at RHIC, and the LHC energies is
rather large, it is obvious that the result of the extrapolation of the 
trend 
of experimental data
from low energies is below the real values of the experimental data  
corresponding to $\sqrt{s_{NN}}$=2.76 TeV and 5.02 TeV LHC energies. 
The second rise  taken at the inflection point in case II (Fig.2-bottom 
plot) corresponds approximately to the rise point,
at the end of the plateau, in case I (Fig.2-upper plot), being around 
$\sqrt{s_{NN}}$=1100 GeV, as deduced from the representation of
$<dN/dy>/S_{\perp}^{geom}=f(\sqrt{s_{NN}})$ in Fig.3. As it will be seen in 
Section V, 
this value agrees with the one corresponding to similar trends evidenced
in the behaviour of the slope of the dependence of the Bjorken energy density times the interaction
time on the entropy density, as a function of the collision energy.
Experimental measurements between  $\sqrt{s_{NN}}$=200 GeV and 2760 GeV, 
missing for the moment, would be of real interest in this context. 

Less influenced by processes such as collective expansion, suppression or 
kinematic cuts, 
the behaviour of such a representation supports the theoretical predictions 
that identify it as a signature of the phase transition
\cite{Bla1, Bla2}. The obvious rise at LHC energies could be the result of
the combined contributions of a higher temperature of the deconfined medium
 and
a higher percolation probability in the regions of high density colour
electromagnetic fields.
\section{$\langle dE_T/dy \rangle/\langle dN/dy \rangle$ - $\langle dN/dy \rangle/S_{\perp}$ correlation for different centralities at RHIC and LHC energies}

\begin{figure} [h]
\includegraphics[width=0.75\linewidth]{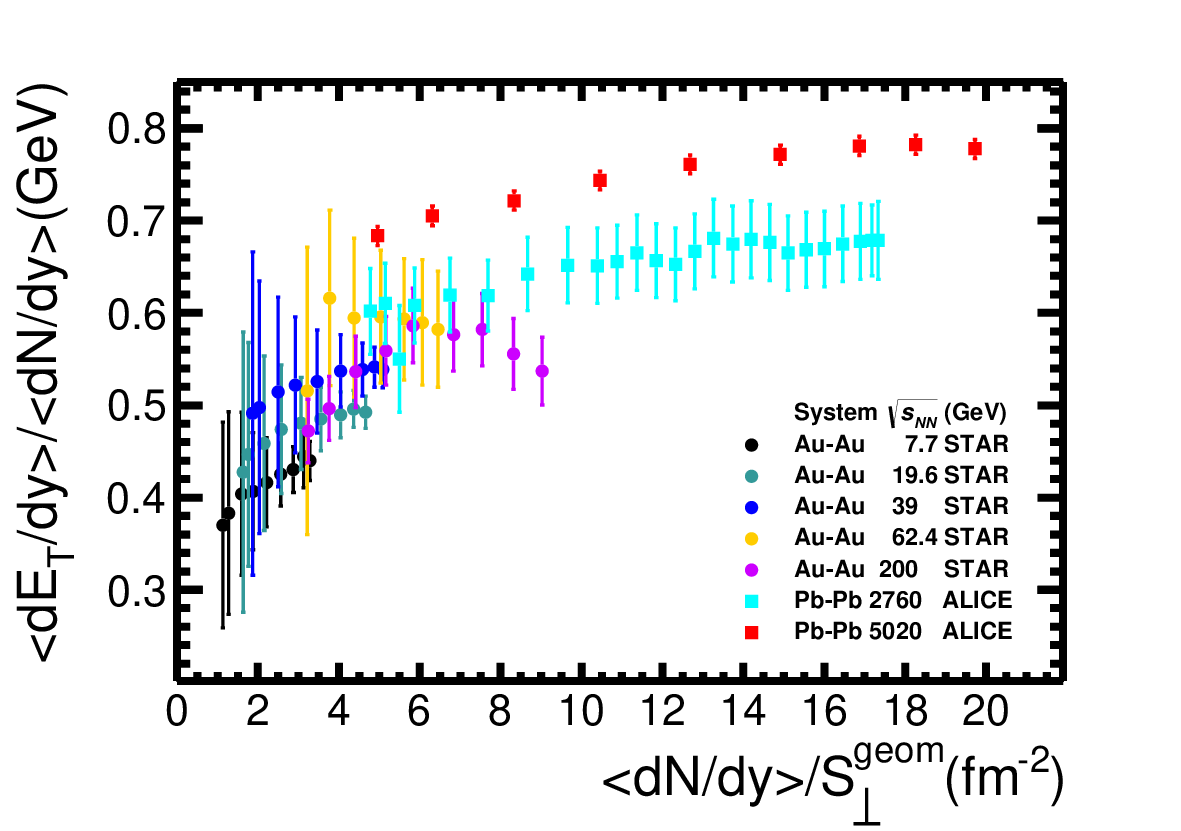}
\includegraphics[width=0.75\linewidth]{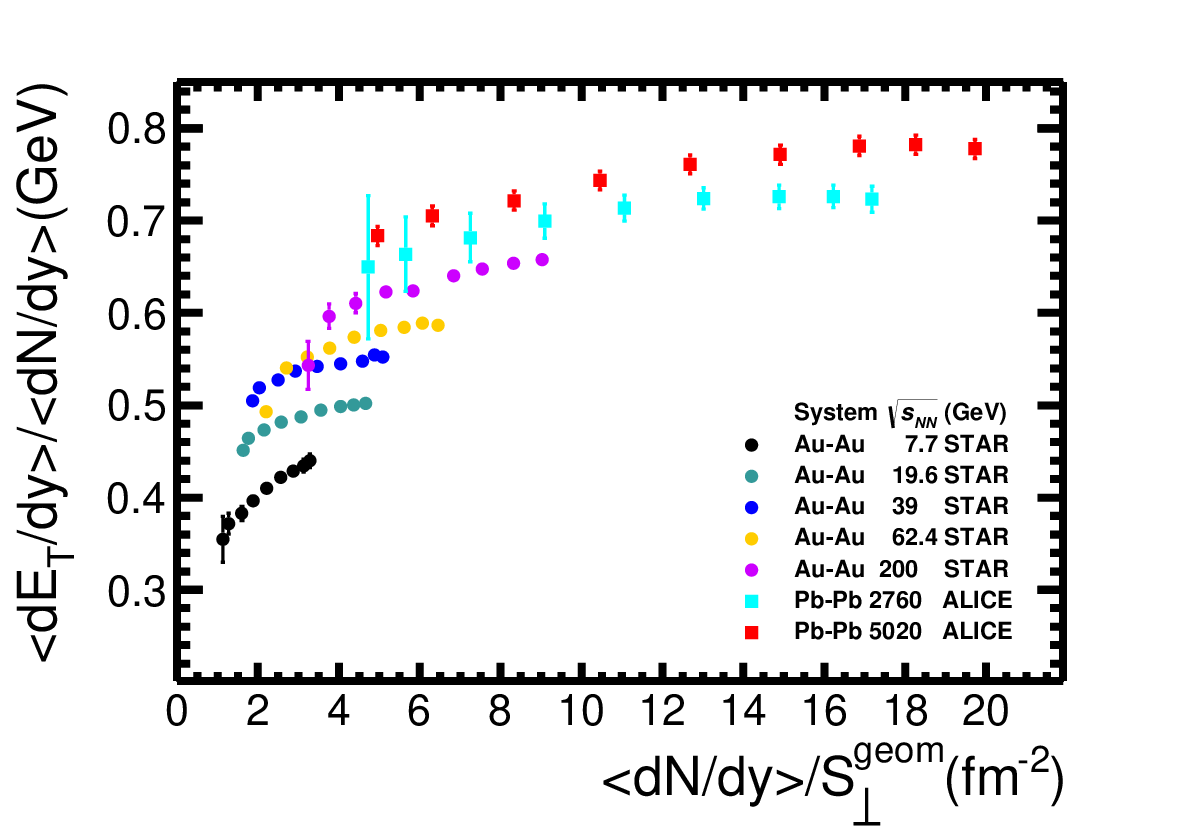}
\caption{The $\langle dE_T/dy\rangle/\langle dN/dy\rangle$ - $\langle dN/dy\rangle/S_{\perp}$ correlation for different collision energies and different centralities. Upper plot: case I, bottom plot: case II.}
\label{fig-4}
\end{figure}

The $\langle dE_T/dy \rangle/\langle dN/dy \rangle$-$\langle 
 dN/dy \rangle/S_{\perp}$ correlation depending on the centrality for several collision energies is shown in Fig.~\ref{fig-4} where the difference between the upper and bottom plots is the same as in Fig.~\ref{fig-2}, explained in the previous chapter.
 \begin{figure}[]
\includegraphics[width=1.0\linewidth]{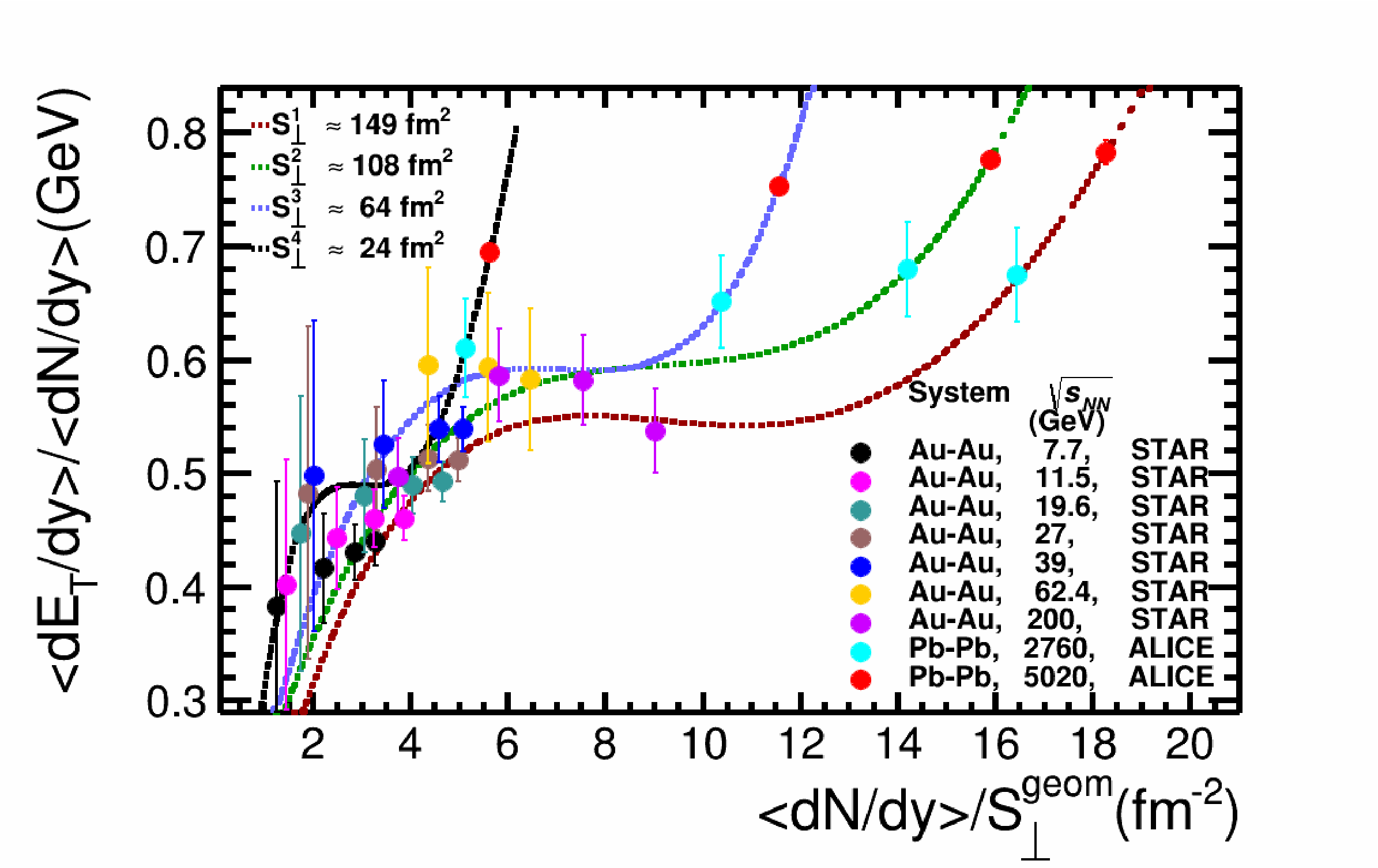}
\includegraphics[width=1.0\linewidth]{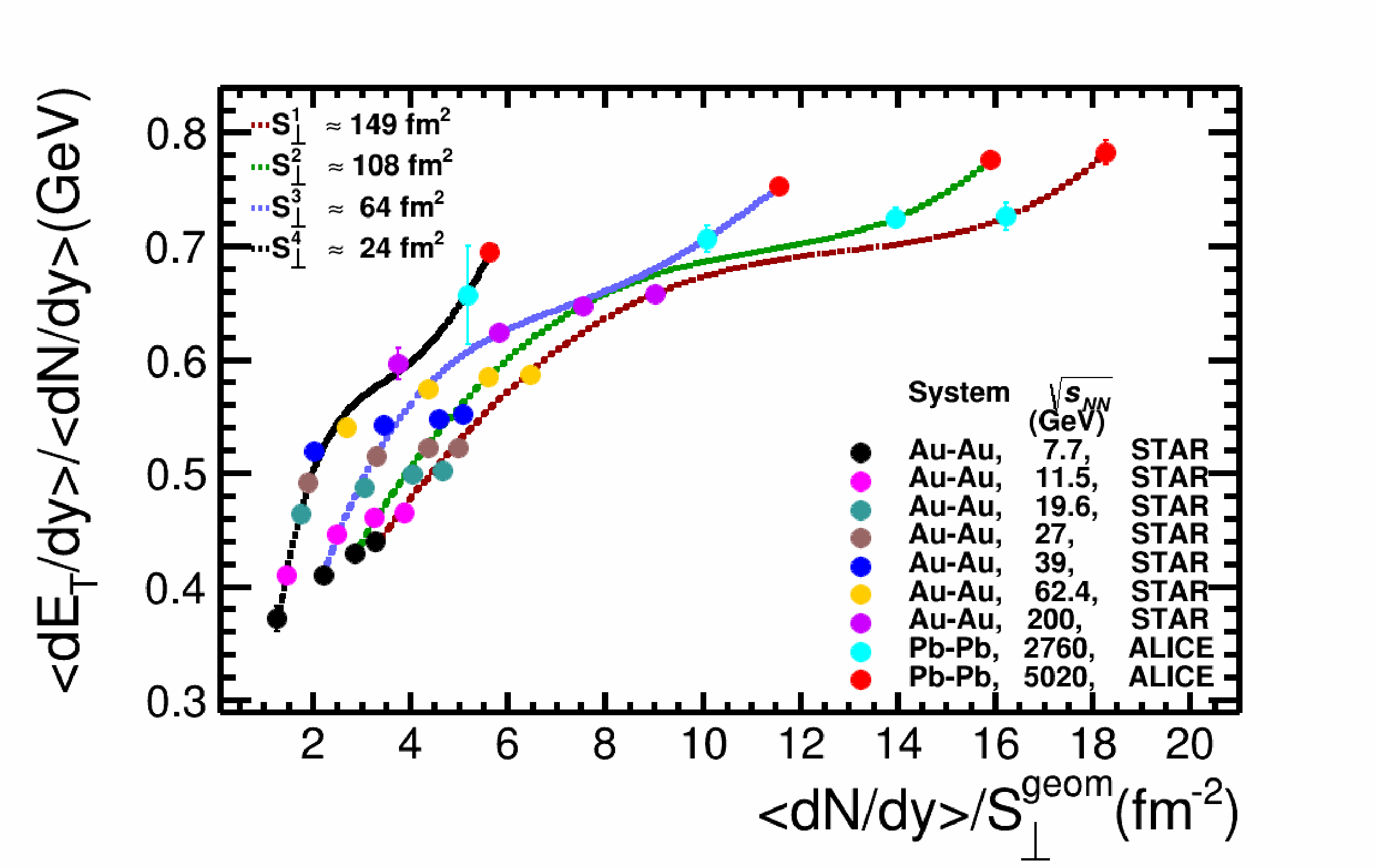}
\caption{The $\langle dE_T/dy \rangle/\langle dN/dy \rangle$ - $\langle dN/dy \rangle/S_{\perp}$ correlation for four values of 
$S_{\perp}^{geom}$ and 
different  collision energies. 
Upper plot: case I, bottom plot: case II. 
The 
 lines represent the results of a
 $4^{th}$ degree polynomial fits of the experimental points.}
\label{fig-5}
\end{figure} 
A close to linear dependence of  $\langle dE_T/dy \rangle/\langle dN/dy \rangle$ as a function of 
 $\langle dN/dy \rangle/S_{\perp}$ can be seen.
  The offsets are 
 different, increasing with the collision energy. 
 The $\langle dE_T/dy \rangle/\langle dN/dy \rangle$-$\langle 
 dN/dy \rangle/S_{\perp}$ correlations for Au-Au at $\sqrt{s_{NN}}$=62.4 GeV \cite{Sahoo1,Sahoo2} and 200 GeV \cite{STAR2} based on calorimetry measurements (case I) 
  do not follow the general trend what  is not observed in case II (Fig.~\ref{fig-4}, bottom).
 In Fig.~\ref{fig-5} are shown the
$\langle dE_T/dy \rangle/\langle dN/dy \rangle$-$\langle dN/dy \rangle/
S_{\perp}$ correlations for the overlap transverse areas $S_{\perp} \approx$ 24, 64, 108 and 149 $fm^2$.   
The  lines represent the result of fitting the corresponding points with a 4th degree polynomial.
A clear dependence on the size of the transverse overlap area is observed.
The rise in 
$\langle dE_T/dy \rangle/\langle dN/dy \rangle$ at RHIC energies becomes
steeper and the range in $\langle dN/dy \rangle/S_{\perp}$, corresponding to 
a close to a plateau trend decreases from central to 
peripheral collisions \cite{Pet1} and is expected to converge towards the values corresponding to minimum bias  pp 
collisions in the same energy range,
once the contribution from nucleons suffering single collisions becomes
predominant.
At each inflection point of the fit lines,
the corresponding $\langle dE_T/dy \rangle/\langle dN/dy\rangle$ value,
related to the temperature of the mixed phase up to the pressure contribution, was estimated.
The representation of these values as a function of $S_{\perp}$ shows a decrease from 
the central collisions to the peripheral ones, as can be seen in Fig.~\ref{fig-6}.

Such a volume dependence was predicted for the Chiral phase transition, the critical 
temperature decreasing towards smaller volume \cite{Bra, Bo}. It is worth mentioning 
that a volume dependence has also been observed in the magnetic phase transition 
\cite{Arn}.
The pressure term in the equation of state could also contribute to the observed trend.
Therefore, theoretical models taking into account all possible contributions such to reproduce the 
experimental trend  are required.

\begin{figure}[]
\includegraphics[width=1.0\linewidth]{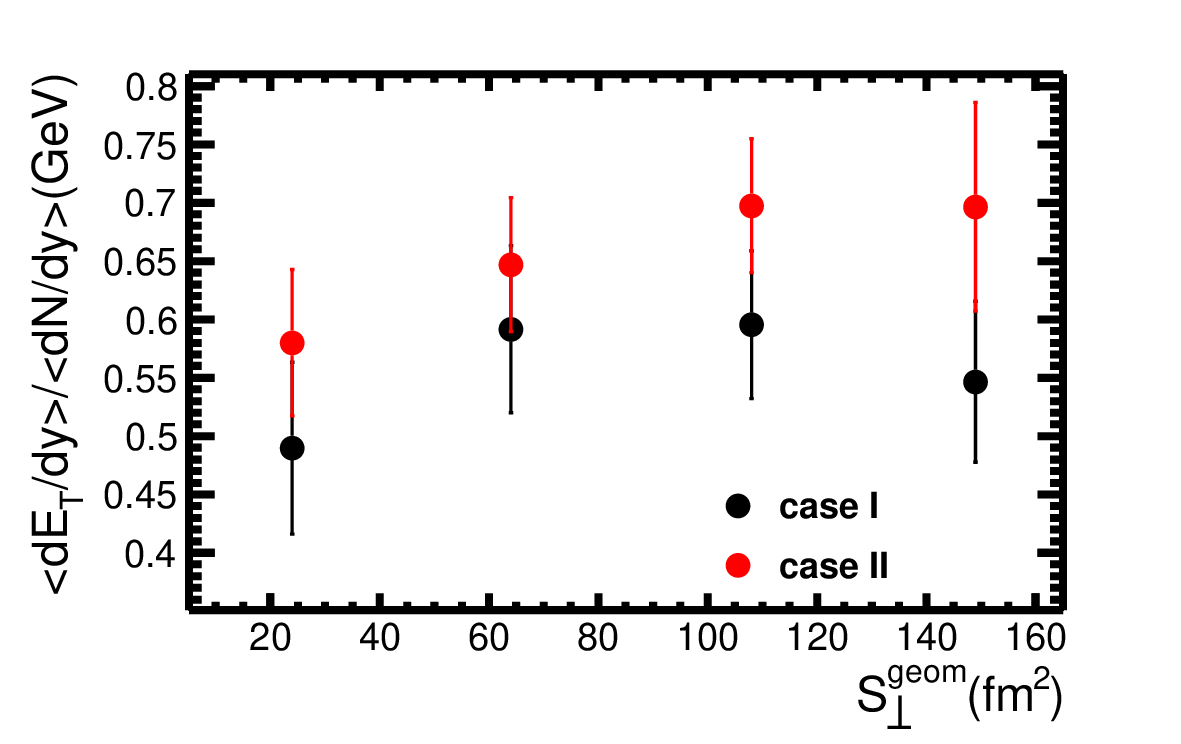}
\caption{$\langle dE_T/dy \rangle/\langle dN/dy \rangle$ calculated at the 
inflection point in the $\langle dE_T/dy \rangle/\langle dN/dy \rangle$-$
\langle dN/dy \rangle/S_{\perp}$ correlation for four values of 
$S_{\perp}^{geom}$ and 
different  collision energies. Black symbols correspond to 
Fig.~\ref{fig-5} - up and the red symbols to 
Fig.~\ref{fig-5} - bottom.}
\label{fig-6}
\end{figure}   

\section{$(\langle dE_T/dy \rangle/\langle dN/dy \rangle)^{core}$-($\langle dN/dy \rangle/S_{\perp})^{core}$ correlation}

In a simple core-corona picture different observables in heavy ion collisions as a 
 function  of centrality and collision energy 
 can be seen as the result of the combination between the contribution 
 coming from the interactions of nucleons undergoing only single collisions (corona) and
 that coming from the multiple interactions of the rest of the nucleons (core).
\begin{figure}
\includegraphics[width=1.0\linewidth]{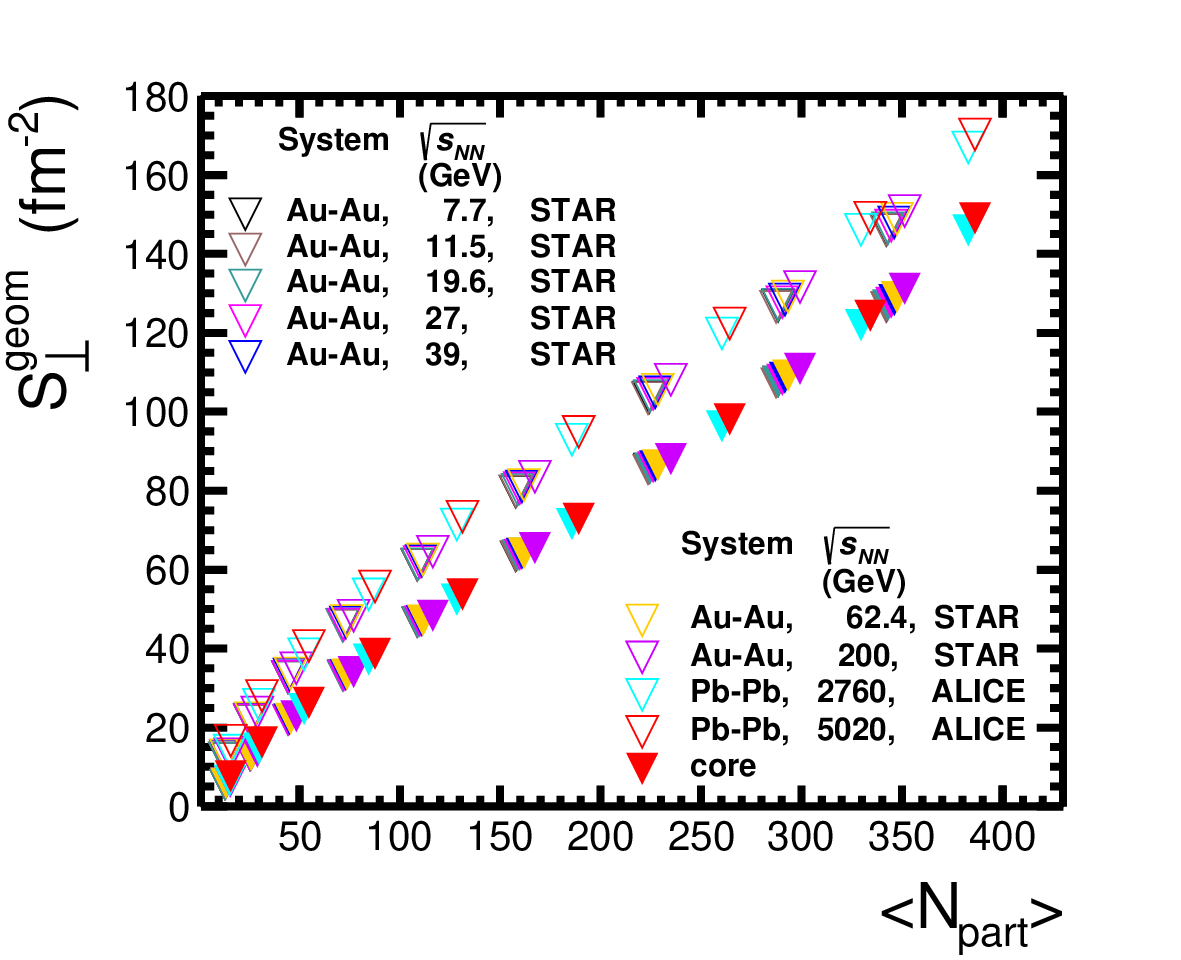}
\caption{The transverse overlap area of the colliding nuclei at different 
energies estimated with the Glauber MC approach
 corresponding to all nucleons ($S_{\perp}^{geom}$) - open symbols and to 
 nucleons 
 suffering more than a single collision 
 ($(S_{\perp}^{geom})^{core}$) - full symbols
as a function of the number of participants,
$\langle N_{part} \rangle$.}
\label{fig-7}
\end{figure}

\begin{figure}[]
\includegraphics[width=1.0\linewidth]{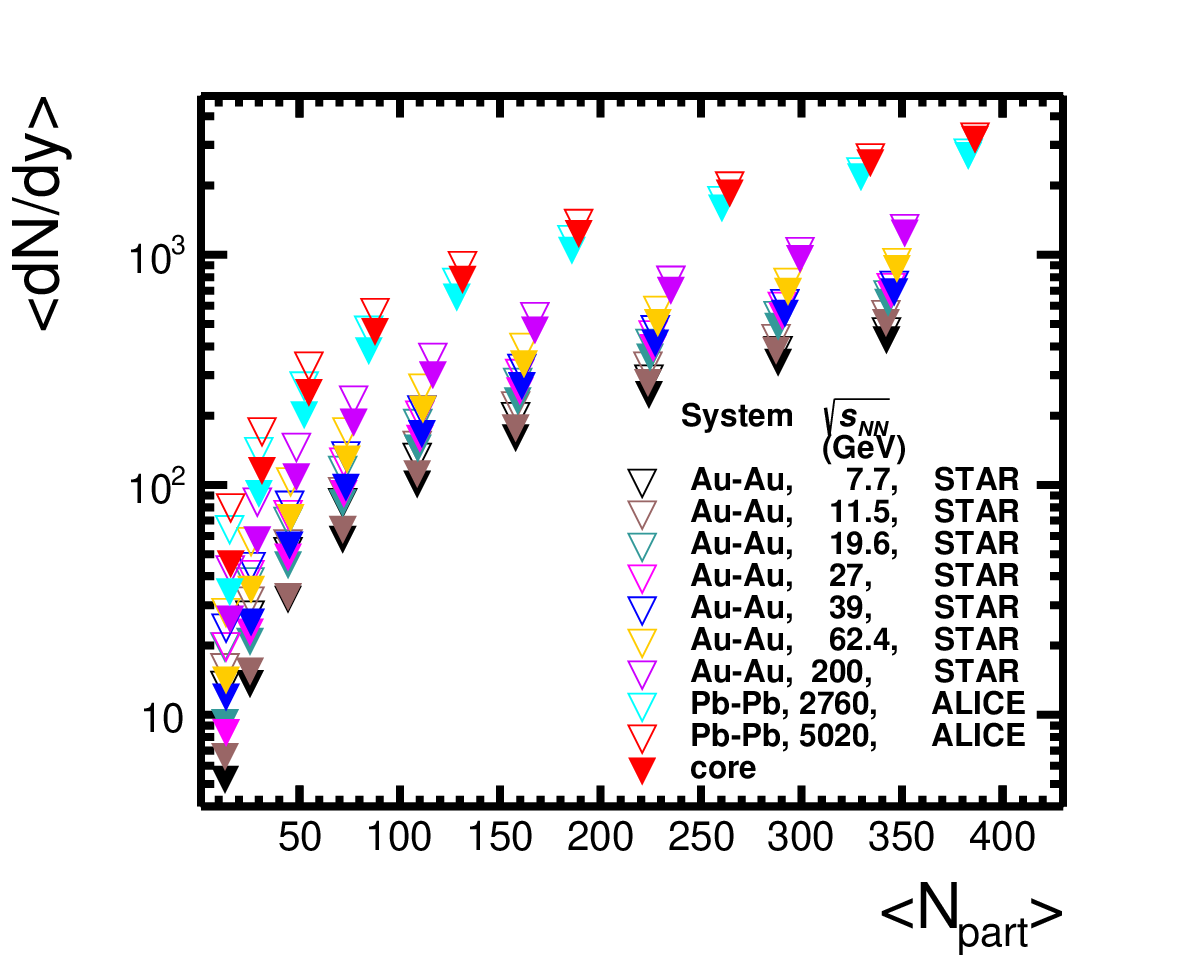}
\caption{$\langle dN/dy\rangle$ estimated by considering only $\pi^{\pm}$, $K^{\pm}$, p and $\bar{p}$ and the 
corresponding neutral 
hadrons
for all A-A collisions at all measured 
energies; open symbols - total, full symbols - core contribution.}
\label{fig-8}
\end{figure}
 \begin{figure}
\includegraphics[width=0.9\linewidth]{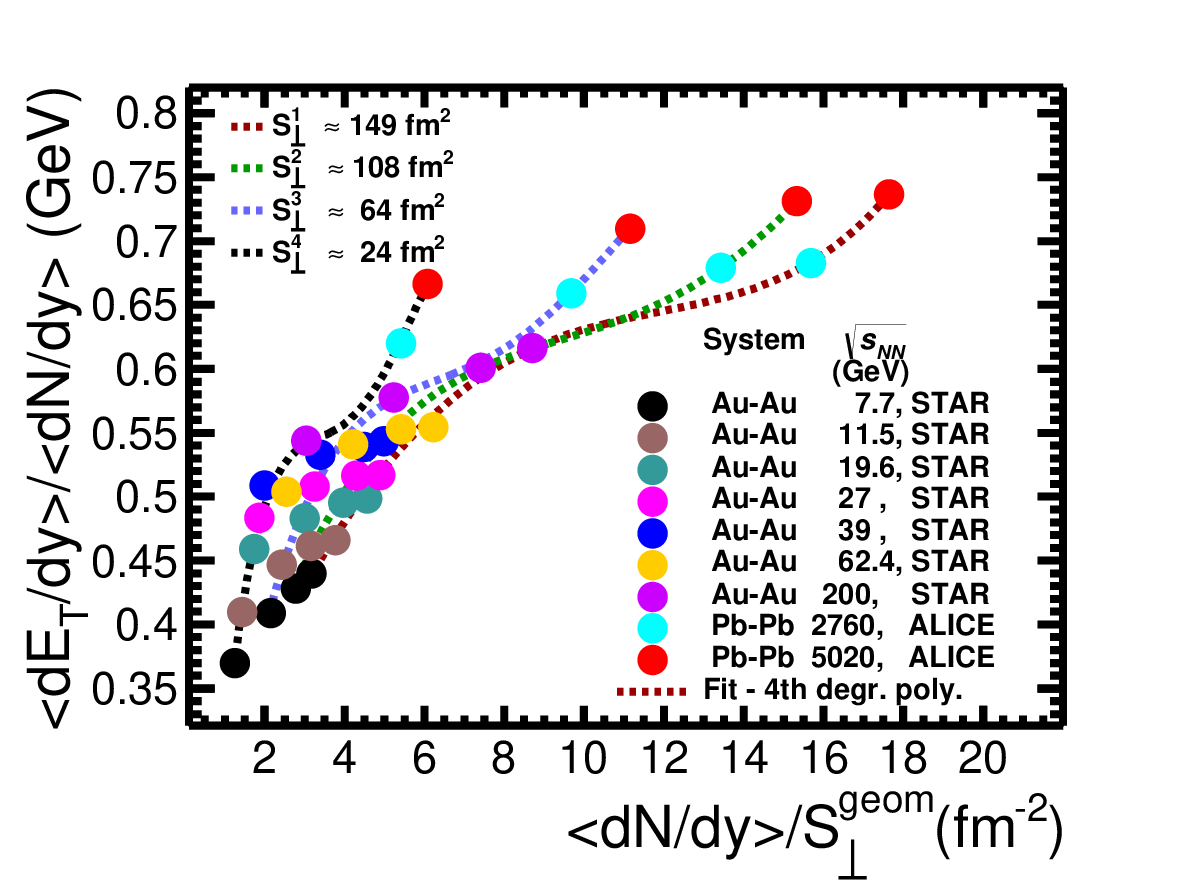}
\includegraphics[width=0.9\linewidth]{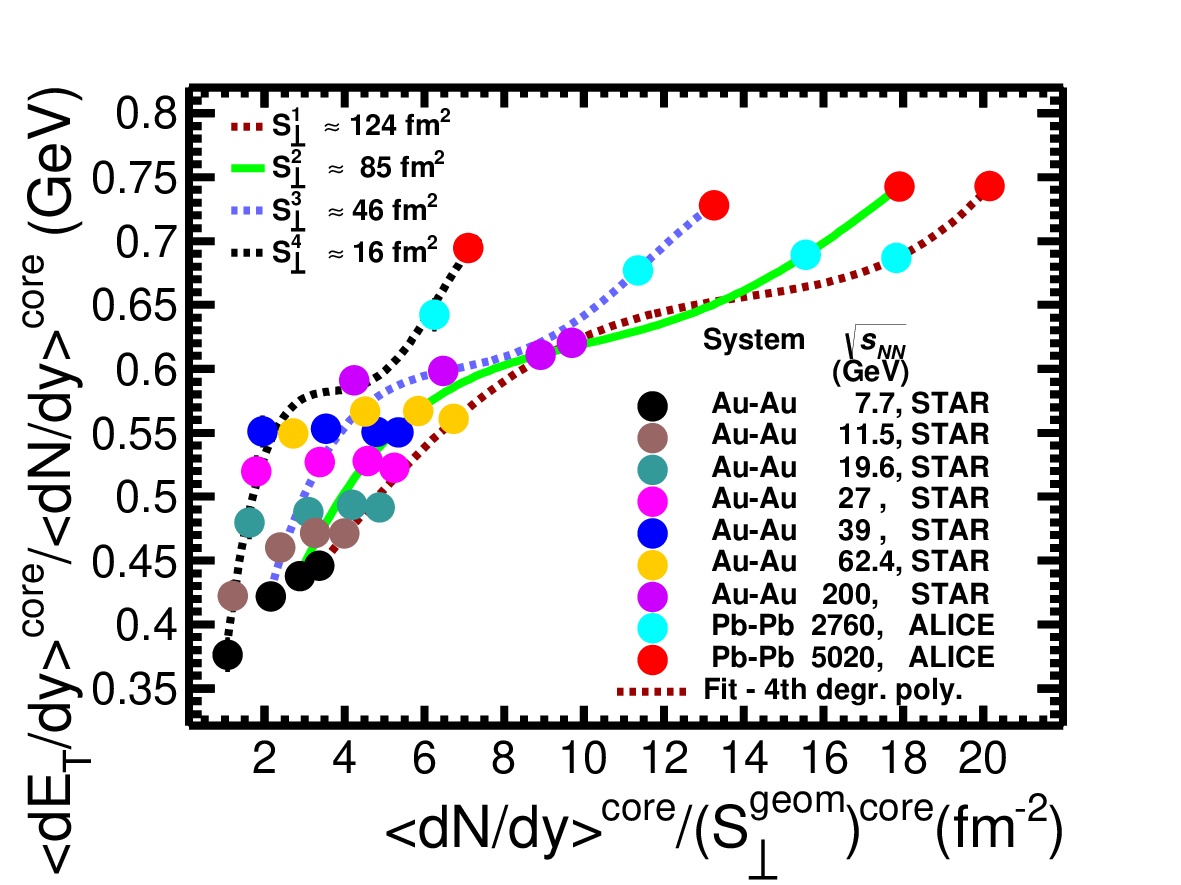}
\caption{Total $\langle dE_T/dy \rangle/\langle dN/dy \rangle)$-$\langle dN/dy \rangle/S_{\perp}$ (upper plot)
and
core $(\langle dE_T/dy \rangle/\langle dN/dy \rangle)^{core}$-$(\langle dN/dy \rangle/S_{\perp})^{core}$ (bottom plot) correlations for four values of the total and core 
transverse overlap area and 
different  collision energies. The lines are the results of the fit with a 4th order polynomial.}
\label{fig-9}
\end{figure}
\begin{figure}[!h]
\includegraphics[width=0.8\linewidth]{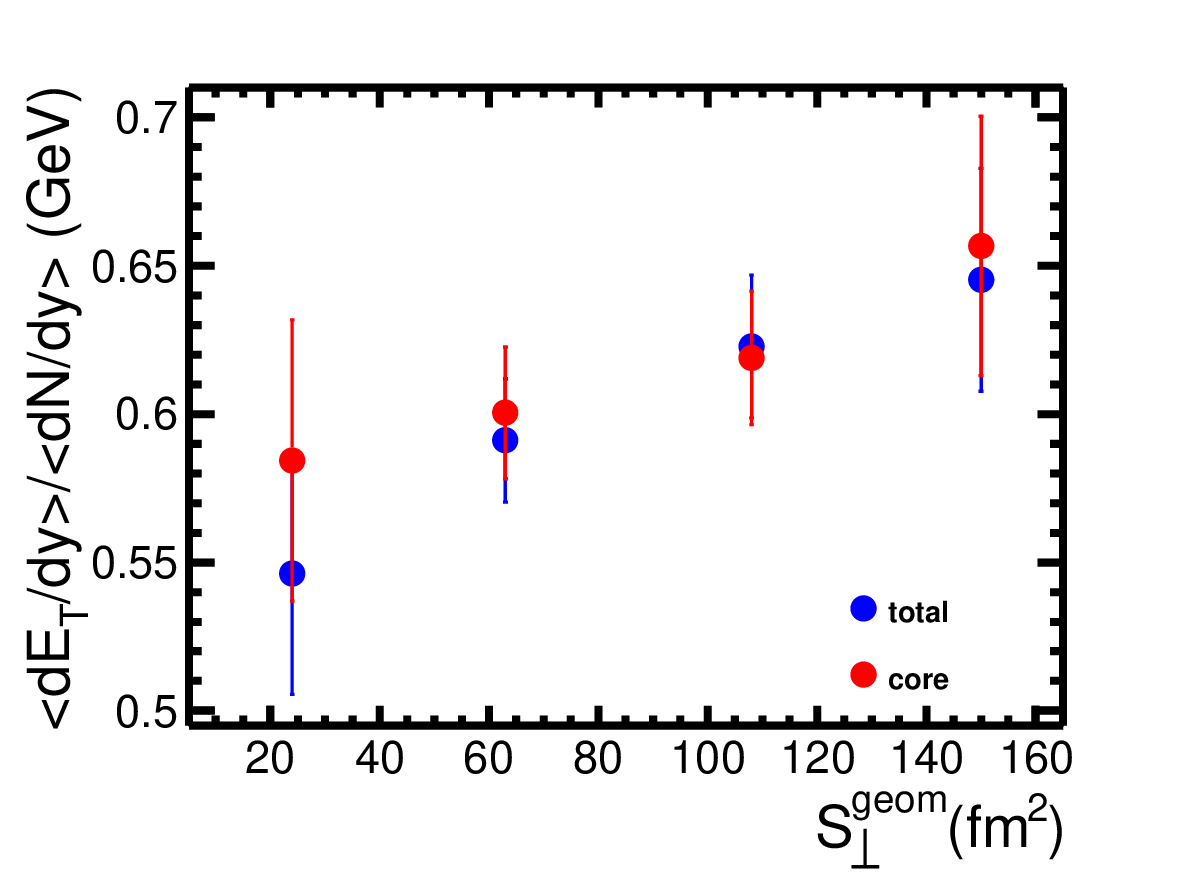}
\caption{$\langle dE_T/dy \rangle/\langle dN/dy \rangle$ calculated at the inflection point in the $\langle dE_T/dy \rangle/\langle dN/dy \rangle$-$\langle dN/dy \rangle/S_{\perp}$ total and core correlations (total - blue symbols, core - red symbols) for four values of the transverse overlap area and 
different  collision energies, as a function of the total transverse overlap area.}
\label{fig-10}
\end{figure}
\noindent
The dependence on
the collision energy and centrality of the core-corona relative weight and its influence on different observables has been studied based on different approaches \cite{Boz, Wer, Bec, Aich1,*Aich2,*Aich3, Boz1, Gem, Pet4, POP1}.
The transverse overlap
areas corresponding to 
interacting nucleons
suffering more than a single collision ($(S_{\perp}^{geom})^{core}$) were estimated within the Glauber Monte-Carlo approach
and compiled in \cite{Pet1}. 
Their $\langle N_{part} \rangle$
dependence is presented in Fig.~\ref{fig-7}.
As far as the $p_T$ spectra are not yet published for all hadrons 
measured in  MB pp collisions, the
studies, whose results are presented in this chapter, were done by 
using  yields and average transverse mass values only for $\pi^{\pm}$, $K^{\pm}$, p and $\bar{p}$ and the 
corresponding neutral 
hadrons in Eqs. \ref{eq4} and \ref{eq5} for both cases, i.e. total and core. 
The core corresponding yield and $\langle m_T \rangle$ for a given species have been estimated 
by subtracting  from the  $p_T$ spectra for A-A at a given centrality the corresponding 
$p_T$ spectra measured in the  MB pp collision at the same energy, weighted with the
number of nucleons suffering only single collisions \cite{POP1}.
$\langle dN/dy\rangle$  total and  corresponding to the core contribution values
are shown in Fig.~\ref{fig-8} as a function of total 
$\langle N_{part}\rangle$. As expected, the corona contribution decreases from 
peripheral towards central collisions. 
The total
$\langle dE_T/dy \rangle/\langle dN/dy \rangle$ - $\langle dN/dy 
\rangle/S_{\perp}$
and core
$(\langle dE_T/dy \rangle/\langle dN/dy \rangle)^{core}$ - ($\langle dN/dy 
\rangle/S_{\perp})^{core}$ correlations  for four corresponding overlap areas are presented in Fig.~\ref{fig-9} upper and bottom 
plots, respectively.  
Although they look
similar, the dependence on the fireball 
transverse area
is more clearly observed especially at lower values of the transverse overlap
area, 
the tendency towards saturation being enhanced for the core contribution. 

 The quantitative difference
 can be followed in Fig.~\ref{fig-10} where are represented 
 the $\langle dE_T/dy \rangle/\langle dN/dy \rangle$ values corresponding to the inflection 
 points of the fit lines for the four values of $S_{\perp}$, total and core. The decrease from central to peripheral
 collisions is similar with the trend presented in the previous chapter where  
 all hadrons were considered. As expected, the main difference between total and core is at peripheral
 collisions. Nevertheless, the decrease of 
 $\langle dE_T/dy \rangle/\langle dN/dy \rangle$ values corresponding to the inflection 
 points from central (larger $S_{\perp}$) to peripheral (lower $S_{\perp}$) is 
  also present when only the core contribution is considered.   

\section{ Bjorken energy density as a function of $\langle dN/dy \rangle/S_{\perp}$}
 \begin{figure} [b]
\includegraphics[width=0.8\linewidth]{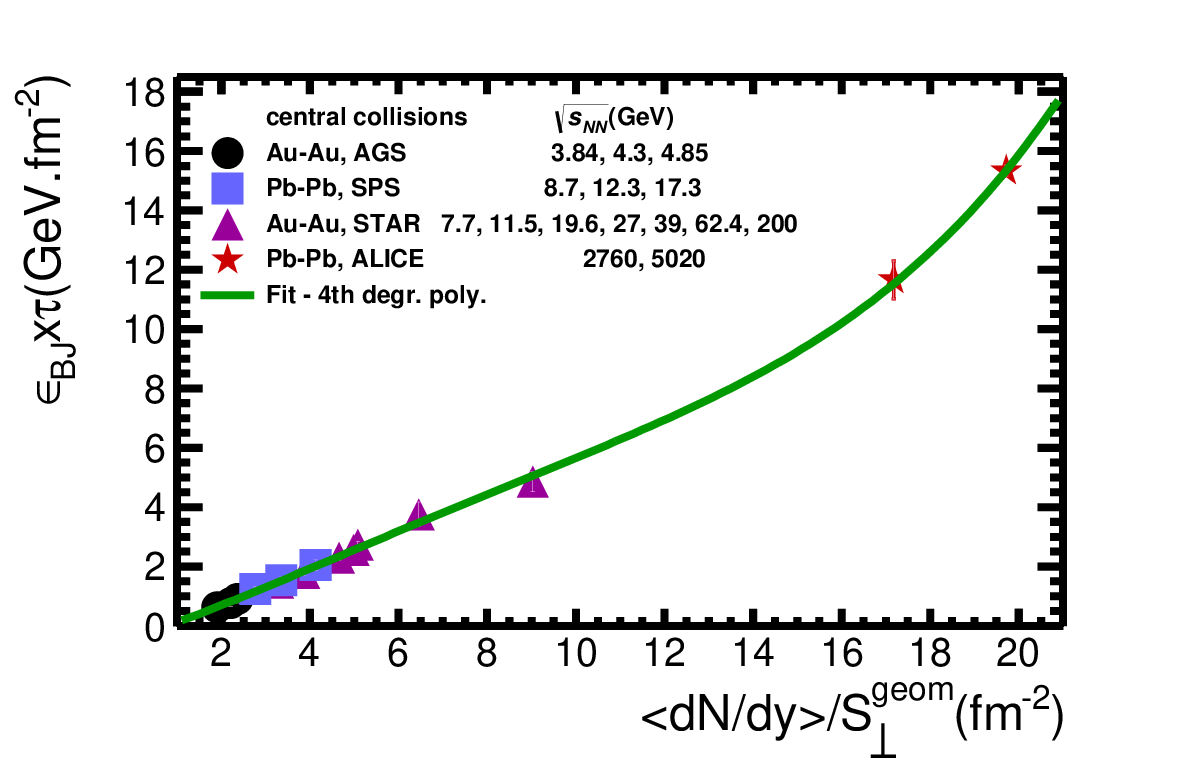}
\includegraphics[width=0.8\linewidth]{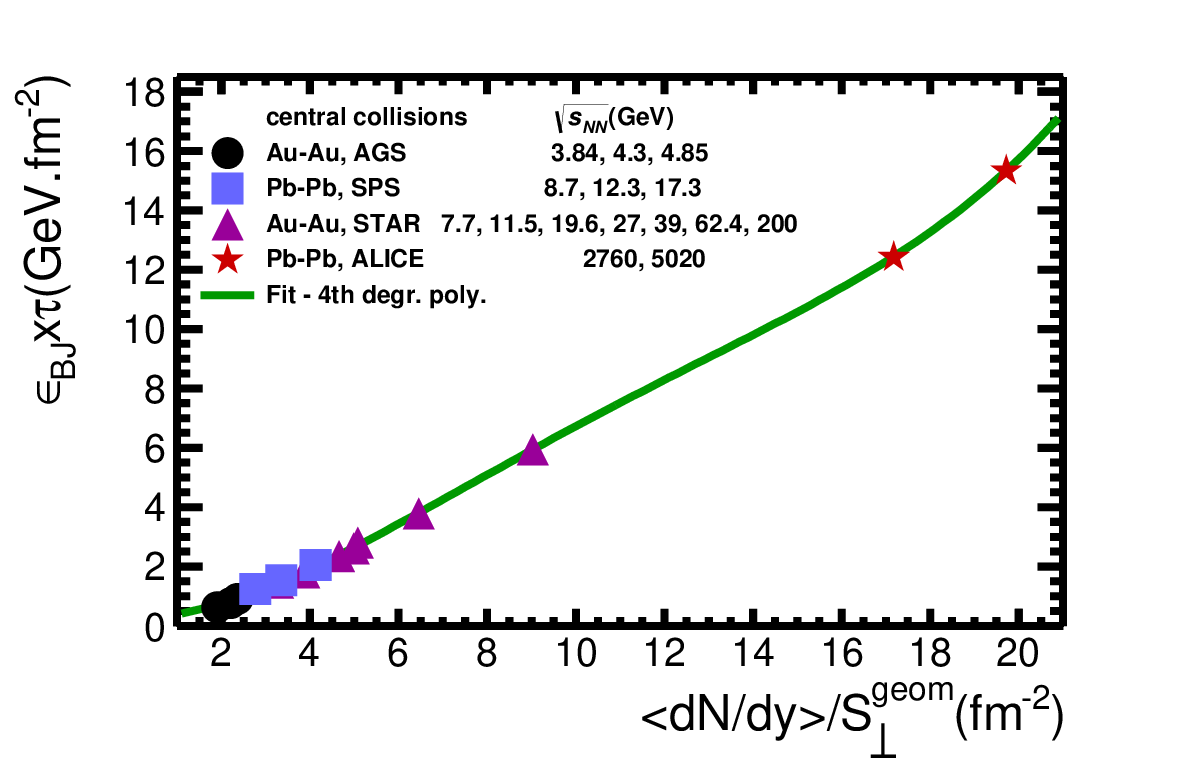}
\caption{The Bjoken energy density times the interaction time as a function of $\langle dN/dy \rangle/S_{\perp}$
for the most central A-A collisions.
Upper plot: case I, bottom plot: case II. 
The green line is the result of a fit with a 4th order polynomial, used to guide the
eye.}
\label{fig-11}
\end{figure} 
\begin{figure}
\includegraphics[width=0.7\linewidth]{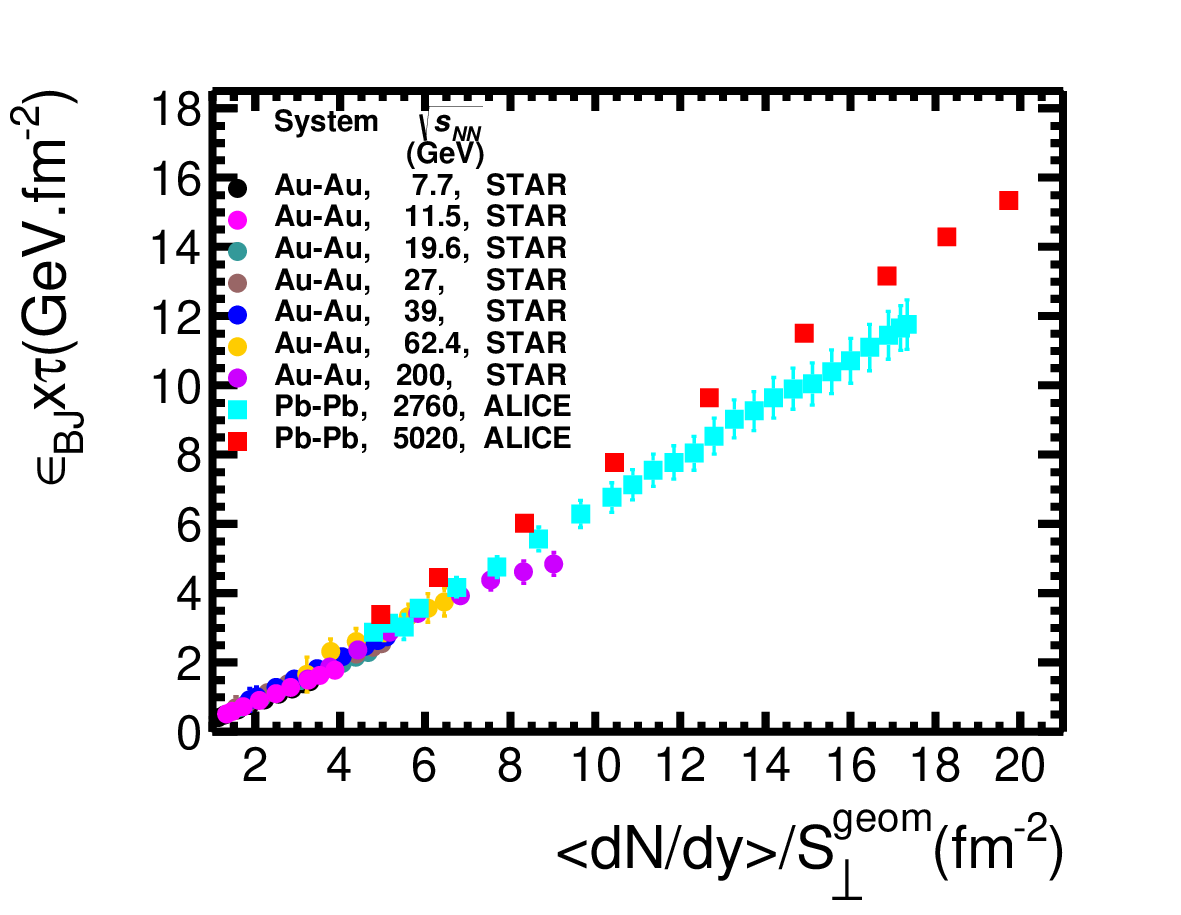}
\includegraphics[width=0.7\linewidth]{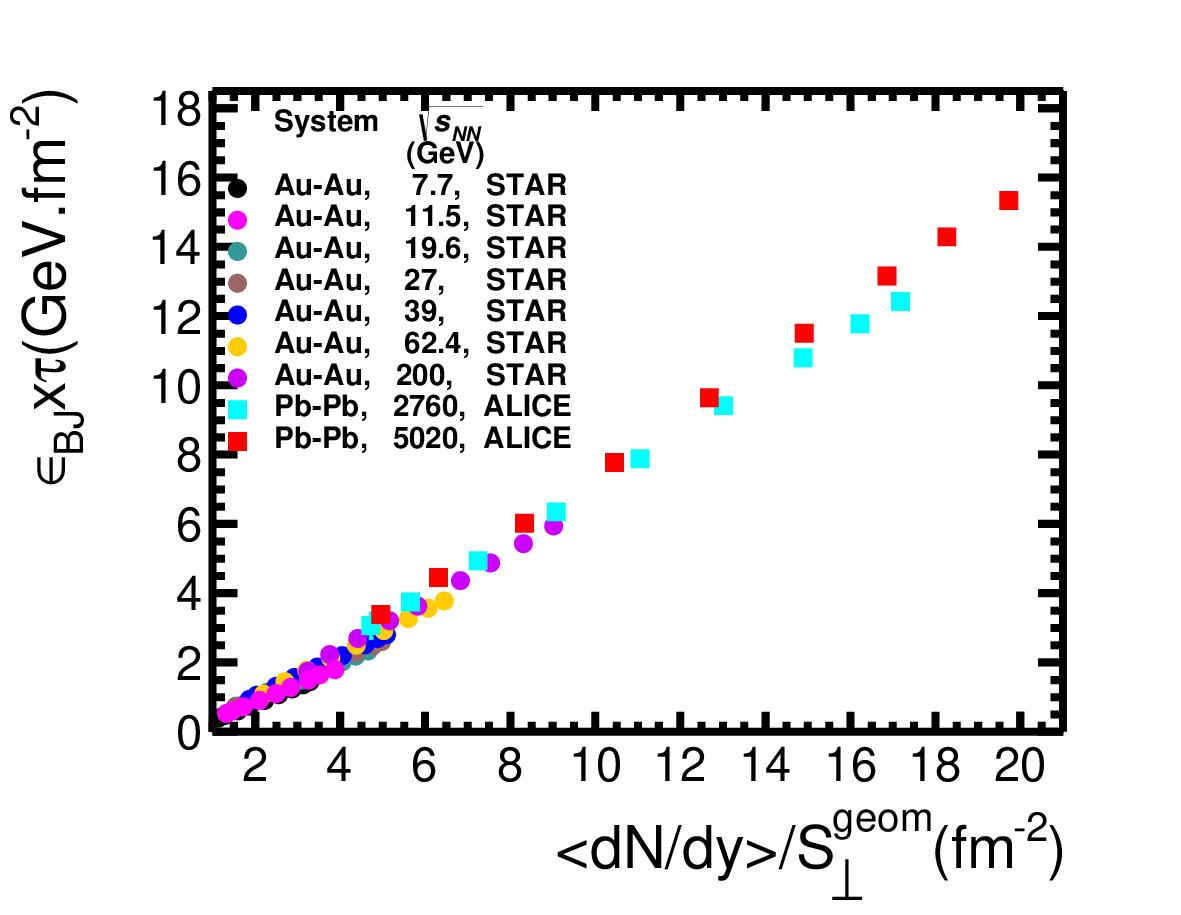}
\caption{The Bjoken energy density times the interaction time as a function of $\langle dN/dy \rangle/S_{\perp}$
for different A-A collision energies and centralities.
The difference between the upper and bottom plots is the same as it is explained
for Fig.~\ref{fig-11}.}
\label{fig-12}
\end{figure}

 The Bjorken energy density times the interaction time 
 ($\epsilon_{Bj}\cdot\tau$), i.e. $\langle dE_T/dy\rangle/S_{\perp}$, as a function of 
 $\langle dN/dy \rangle/S_{\perp}$ for the most central A-A collisions considered in the 
 previous sections is presented in Fig.~\ref{fig-11}.
 The upper and bottom plots were obtained in case I and II as discussed in 
 Chapters II and III.

 The $\epsilon_{Bj}\cdot\tau$
 from AGS, SPS and RHIC energies has a linear dependence on the
 particle multiplicity per unit rapidity and unit transverse overlap 
 area $\langle dN/dy \rangle/S_{\perp}$. 
 The extrapolation to the $\langle dN/dy \rangle/S_{\perp}$ values
 corresponding to the LHC energies falls  below the experimental
 results. In Fig.~\ref{fig-12} are represented the 
 $\epsilon_{Bj}\cdot\tau$ values for different centralities and collision
 energies. The upper and bottom plots were obtained for case I and II as explained in 
 Sections II and III.

\begin{figure}
\includegraphics[width=0.8\linewidth]{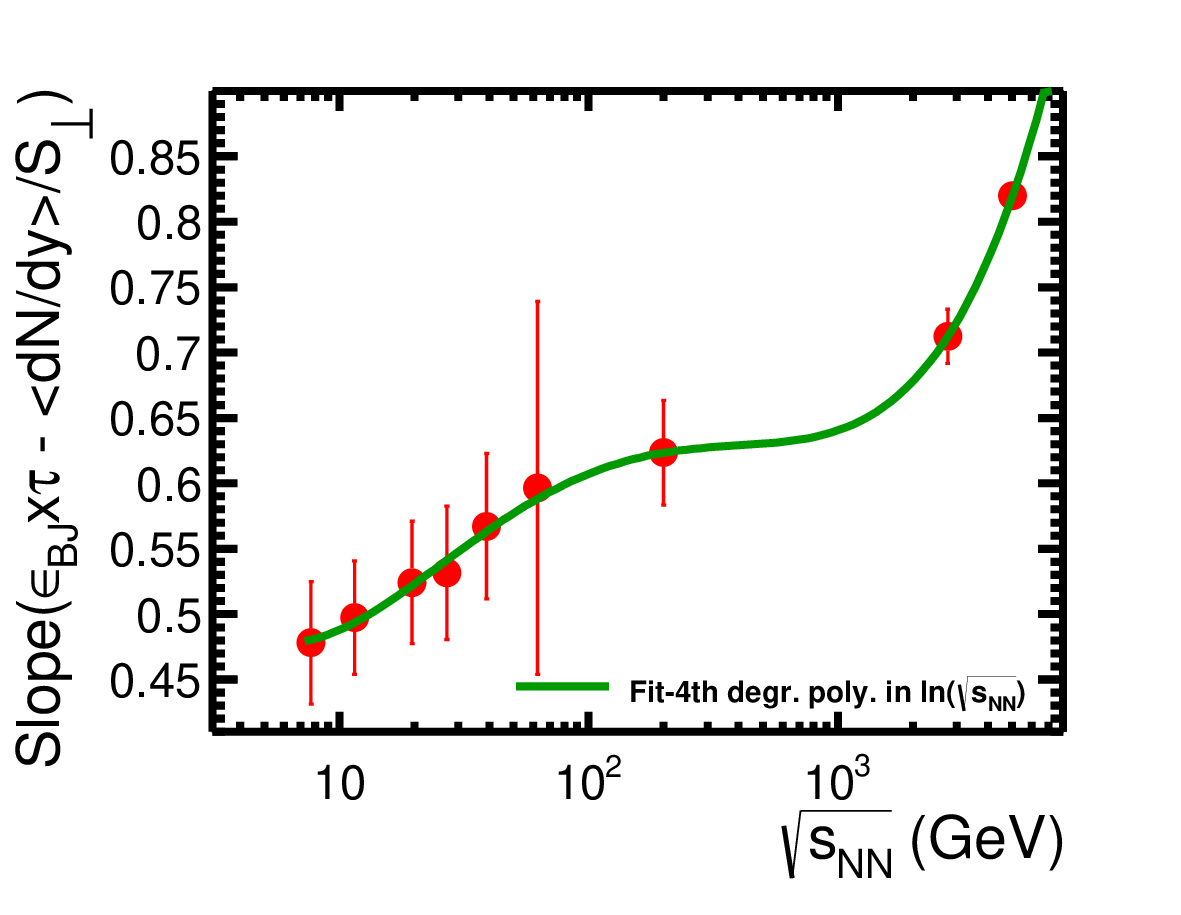}
\includegraphics[width=0.8\linewidth]{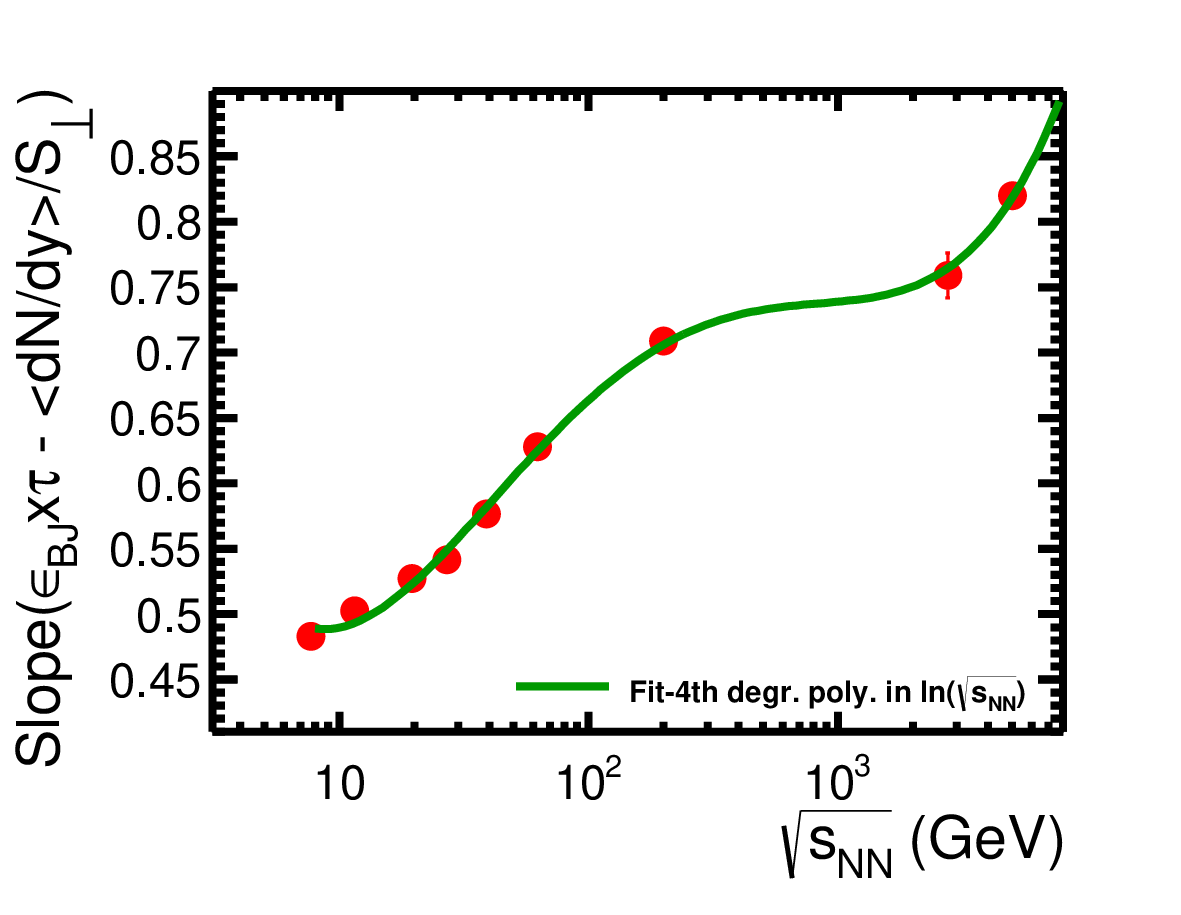}
\caption{The slope of the $\epsilon_{Bj}\cdot\tau$ linear dependence on 
$\langle dN/dy \rangle/S_{\perp}$ as a function of the collision energy. The line is the result
of the fit with a 4th order polynomial in ln($\sqrt{s_{NN}}$) and is used to guide the eye. Upper plot: case I, bottom plot: case II.}
\label{fig-13}
\end{figure}

For each collision energy 
$\epsilon_{Bj}\cdot\tau$ shows a linear dependence on  
$\langle dN/dy \rangle/S_{\perp}$. The values of the slope of this dependence, related to the temperature of the system,
resulting from a linear fit 
of this correlation for each system and energy,
as a function of collision energy are presented in Fig.~\ref{fig-13}. 

The slope values increase from AGS to the highest RHIC energy. The values for LHC 
energies are larger and not in-line with what should be expected from a simple 
extrapolation of the trend observed at lower collision energies, thus suggesting a
saturation or an inflection point within the energy gap between RHIC and LHC energies. 
Following the considerations presented in Section II, the physics behind the increasing
trend at LHC energies could be related to the large increase of the gluon density, 
therefore increasing the percolation probability, the ropes decay being at the origin 
of increased average transverse momentum.   

\section{Similar studies for $pp$ MB for different collision energies 
and as a 
function of $\langle dN/dy \rangle/S_{\perp}$ at LHC energies}

\begin{figure}[h]
\includegraphics[width=0.8\linewidth]{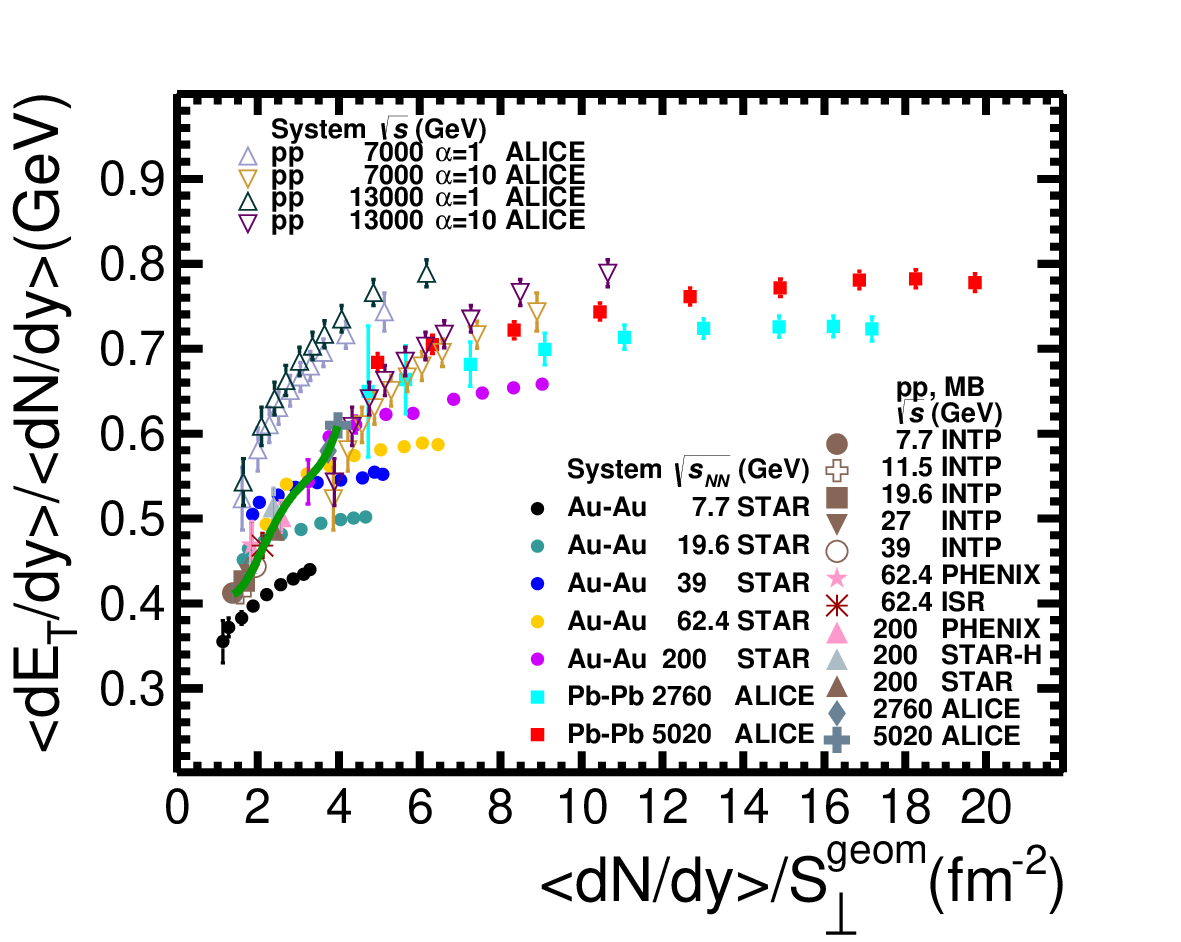}
\caption{The $\langle dE_T/dy \rangle/\langle dN/dy \rangle)$-$\langle dN/dy 
 \rangle/S_{\perp}$ correlation for A-A collisions, case II,  pp MB at the same energies (the green line, the result of a 4th order polynomial, guides the eye)
   and pp as a function of charged particle multiplicity
 at $\sqrt{s}$ = 7 and 13 TeV.
 }
\label{fig-14}
\end{figure}

The experimental data measured in pp MB collisions
\cite{STAR1,ALICE5,PHENIX3,STAR-H1,STAR-H2,ALICE12}  at
$\sqrt{s}$ = 62.4 GeV, 200 GeV, 2.76 TeV and 5.02 TeV were used for
obtaining 
the $\langle dE_T/dy \rangle/\langle dN/dy \rangle)$-$\langle dN/dy \rangle/S_{\perp}$ 
correlation considering only  $\pi^{\pm}$, $K^{\pm}$, p and $\bar{p}$ and the 
corresponding neutral 
hadrons in Eqs. \ref{eq4} and \ref{eq5}. For the BES energies the $p_{T}$ spectra of the same particles, obtained with an interpolation procedure (INTP) from 
existing data in this range of energies \cite{ISR} as explained in \cite{POP1},  were used.  The transverse overlap area was estimated using the Glauber MC approach at subnucleonic level \cite{Glis2}. In Fig.~\ref{fig-14} this correlation is presented on top of the results  
presented in Fig.~\ref{fig-4}, bottom plot.  
It is seen that the correlation 
in A-A collisions at different collision energies converges 
for very peripheral collisions, towards the one
corresponding to the pp MB collisions.
This is expected as far as towards 
peripheral A-A collisions the majority of interacting nucleons suffer 
only single nucleon-nucleon interactions.   
 
At LHC energies, studies of pp collisions up to very high charged 
particle multiplicities, have shown similarities between 
pp and Pb-Pb in terms 
of the behaviour of different observables, like 
near-side long range 
pseudorapidity correlations
\cite{CMS1},  
the ($\langle\beta_T\rangle$ - $T_{kin}^{fo}$) correlation as a function of
charged particle multiplicity \cite{Cristi1},
azimuthal angular correlations \cite{CMS2},
geometrical scaling
\cite{Pet1, Pet2, Pet3}, etc.
Experimental data on transverse momentum spectra for light flavors as a function of the average charged particle 
multiplicity at LHC energies were published in \cite{ALICE9,ALICE10,ALICE11}.
The transverse overlap area for pp collisions, $S_{\perp}^{pp}$=$\pi$
$r_{max}^2$ as a function of charged particle multiplicity  was
estimated using the result of \cite{Bzdak}, computed in the IP-Glasma 
model, 
$r_{max}$ being the maximal radius for which the energy density of the 
Yang-Mills fields is above 
$\varepsilon=\alpha\Lambda_{QCD}^4$ ($\alpha\in{[1,10]}$).
The model results were fitted with: 
\begin{equation}
f_{pp}=
\left\{
\begin{array}{rl}
0.387+0.0335x+0.274x^2-0.0542x^3 & \mbox{if $x<3.4$}\\
1.538                            & \mbox{if $x\geq$ 3.4} 
\end{array}
\right.
\end{equation}
Using the same recipe we fitted the $r_{max}$ values from Ref. \cite{Bzdak} for $\alpha$=10 with the following expression:
\begin{equation}
f_{pp}=
\left\{
\begin{array}{rl}
-0.18+0.3976x+0.095x^2-0.028x^3 & \mbox{if $x<3.4$}\\
1.17                            & \mbox{if $x\geq$ 3.4} 
\end{array}
\right.
\end{equation}
where x=$(dN/dy)^{1/3}$ and $r_{max}$=1fm$\cdot f_{pp}(x)$. 
\begin{figure}[h]
\includegraphics[width=0.8\linewidth]{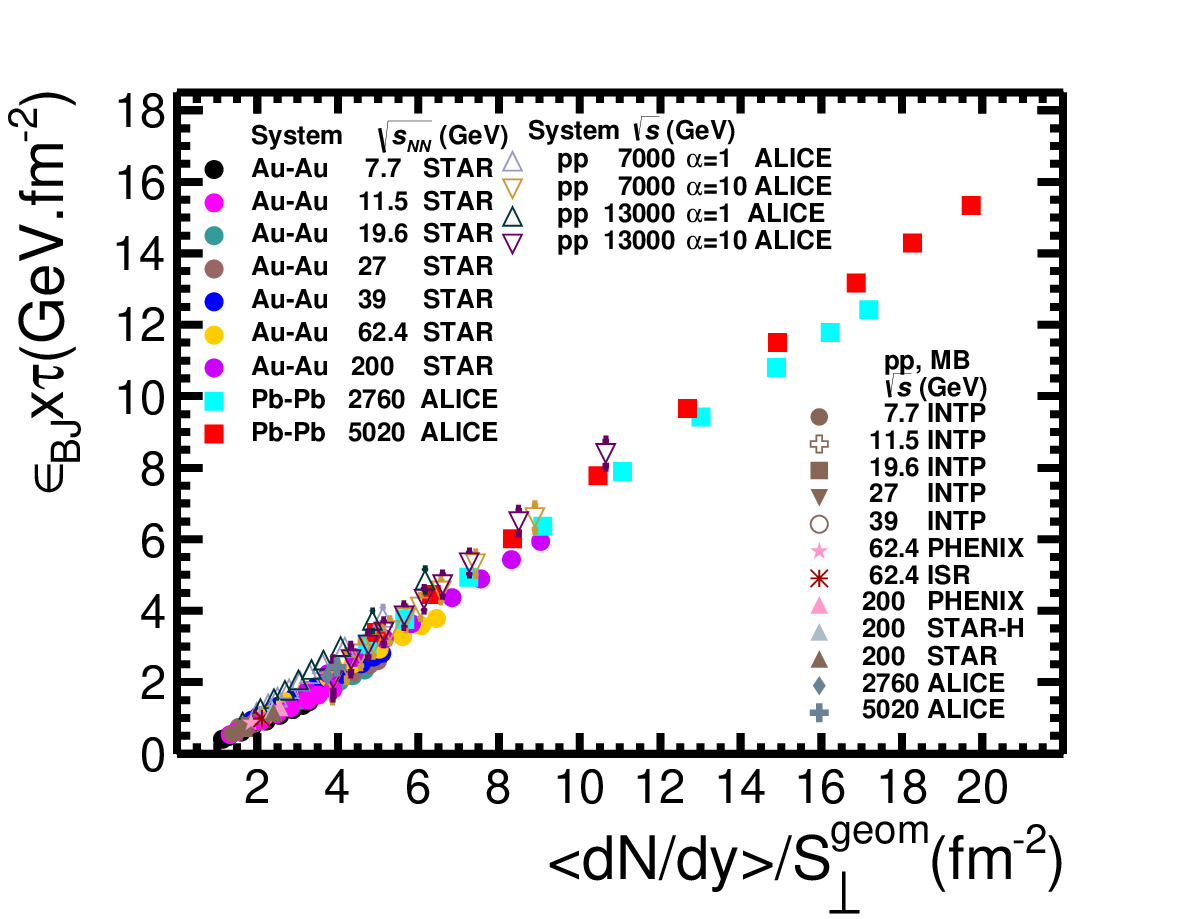}
\caption{The same type of figure as Fig.~\ref{fig-14}, but for the Bjorken energy density times the interaction time as a function of $\langle dN/dy 
 \rangle/S_{\perp}$. }
\label{fig-15}
\end{figure}

The transverse energy per unit  rapidity has been estimated using the same 
procedure as the one used for A-A collisions at LHC explained in 
Chapter II. 
 The result in terms of 
 $\langle dE_T/dy \rangle/\langle dN/dy \rangle$ - $\langle dN/dy 
 \rangle/S_{\perp}$ correlations for pp collisions at 
 $\sqrt{s}$=7 TeV  and 13 TeV is presented in
 Fig.~\ref{fig-14} for $\alpha$=10 and $\alpha$=1.
 The results corresponding to pp collision at 
 $\sqrt{s}$=13 TeV were obtained within the assumption that the above 
 parameterization of 
 the transverse overlap area does not  change significantly. 
 It is seen that the trend of the
 $\langle dE_T/dy \rangle/\langle dN/dy \rangle$ - $\langle dN/dy 
 \rangle/S_{\perp}$ correlation for pp collisions at the LHC energies
 follows the one of Pb-Pb at $\sqrt{s_{NN}}$=5.02 TeV, the quantitative agreement 
 being good for $\alpha$=10 within the same 
 $\langle dN/dy\rangle/S_{\perp}$ range. For the same values of
 $\langle dN/dy\rangle/S_{\perp}$ slightly larger values of 
 $\langle dE_T/dy \rangle/\langle dN/dy \rangle$ in pp 
 at $\sqrt{s}$=7 TeV relative to Pb-Pb at $\sqrt{s_{NN}}$=5.02 TeV and
 in pp at $\sqrt{s}$=13 TeV relative to $\sqrt{s}$=7 TeV are seen.
 As mentioned in the previous chapters, this is in line with
 the expectations based
 on the string percolation approach. The Bjorken energy
 density times the interaction time values for pp collisions at the two LHC energies as a function of 
 $\langle dN/dy\rangle/S_{\perp}$ presented in Fig.~\ref{fig-15}, using the same 
 symbols as in Fig.~\ref{fig-14}, 
 are in very good agreement with the ones corresponding
 to Pb-Pb collision at $\sqrt{s_{NN}}$=5.02 TeV.
For pp collisions the slopes of $\epsilon_{Bj}\cdot\tau$ as a function of  $\langle dN/dy\rangle/S_{\perp}$
are 0.89$\pm$0.09 at $\sqrt{s}$ = 7 TeV and 0.92$\pm$0.07 at $\sqrt{s}$ = 13 TeV, respectively, for $\alpha$ =10,
a bit  larger values than that corresponding to the Pb-Pb collision at $\sqrt{s_{NN}}$ = 5.02 TeV.

\section{Conclusions} 

The 
behaviour of $\langle p_T \rangle/\sqrt{\langle dN/dy\rangle/S_{\perp}} $
as a function of the collision energy for a given centrality or
as a function of centrality for a given collision energy supports the 
predictions
of CGC and percolation based approaches.
The dependence of the ratio of the energy density
to the entropy density as a function of entropy density at 
different collision centralities for A-A collisions from AGS, SPS, RHIC 
and LHC energies reveals a tendency towards saturation starting with the
largest RHIC energies and a steep rise at the LHC energies. 
A clear dependence of this behaviour on the size of the transverse overlap area
is highlighted.
The increase in 
$\langle dE_T/dy \rangle/\langle dN/dy \rangle$ at RHIC energies becomes
steeper and the  $\langle dN/dy \rangle/S_{\perp}$ range, which corresponds 
to a  trend close to a plateau, decreases from central to 
peripheral collisions, converging towards the behavior corresponding
to MB pp collisions. 

In a core-corona scenario, the corona contribution to the observed trends is removed and 
similar correlations are presented for the core in the case of the identified charged hadrons 
and their corresponding neutral particles.

The values of the
energy density to the entropy density for the same entropy density
increase with the collision energy, supporting the expectation based on 
the string percolation approach. 

The Bjorken energy density times the interaction time has a linear 
dependence on the
particle multiplicity per unit rapidity and unit transverse overlap area. The linear 
dependence also holds
for the LHC energies, its slope increasing significantly at these energies which is 
an expected trend in the string percolation approach.

The observed trend of the slopes of the $\epsilon_{Bj}\cdot\tau$ dependence on the entropy 
density $\langle dN/dy\rangle/S_{\perp}$,
as a function of the collision energy, is similar with the ones evidenced in the 
$\langle dE_T/dy \rangle/\langle dN/dy \rangle$-$\langle dN/dy\rangle/S_{\perp}$ 
correlations.

The 
$\langle dE_T/dy \rangle/\langle dN/dy \rangle$-$\langle dN/dy\rangle/S_{\perp}$ and 
$\epsilon_{Bj}\cdot\tau$-$\langle dN/dy\rangle/S_{\perp}$ correlations for pp collisions at 
$\sqrt{s}$=7 TeV and 13 TeV follow qualitatively the ones corresponding to
Pb-Pb collisions at $\sqrt{s_{NN}}$=5.02 TeV. 
Within the error bars, 
there is
also a quantitative agreement if for $S_{\perp}$ for the pp collision as a 
function of particle density only the region of 
overlap zone characterized by an energy density larger than 
$\approx$ 2 GeV/$fm^3$ is considered. 
For pp collisions the slopes corresponding to the $\epsilon_{Bj}\cdot\tau$-$\langle dN/dy\rangle/S_{\perp}$ correlation have 
a bit  larger values than that corresponding to the Pb-Pb collision at $\sqrt{s_{NN}}$ = 5.02 TeV.

\section*{ACKNOWLEDGMENTS}

This work was carried out under the contracts sponsored by the Ministry of 
Research, Innovation and Digitization:  RONIPALICE-07/03.01.2022 (via IFA 
Coordinating Agency)
 and PN-19 06 01 03.	
\nocite{*}
\bibliography{phase_transition}% Produces the bibliography via BibTeX.
\end{document}